\documentclass[apj]{emulateapj}

\usepackage{natbib}

\usepackage{graphicx}
\usepackage{epstopdf}
\usepackage{natbib}
\bibpunct{(}{)}{;}{a}{}{,}
\usepackage{amssymb,,amsmath}
\usepackage{longtable}
\usepackage{multirow}
\usepackage{txfonts}
\usepackage[switch]{lineno}
\usepackage{ifthen}
\usepackage[breaklinks,colorlinks,citecolor=blue]{hyperref}

\usepackage{morefloats}
\usepackage{natbib}
\begin{document}


\newcommand{\nbb}[0]{{0\nu}\beta\beta}
\newcommand{\dlmm}[0]{d^l_{mm^\prime}(\alpha)}
\newcommand{\rs}[0]{r_{\text{s}}}
\newcommand{\vn}[0]{v_{\text{NW}}}
\newcommand{\vs}[0]{v_{\text{sol}}}
\newcommand{\vesc}[0]{v_{\text{esc}}}
\newcommand{\re}[0]{R_{\text{E}}}
\newcommand{\rd}[0]{R_{\text{D}}}
\newcommand{\me}[0]{M_{\text{E}}}
\newcommand{\Cmm}[0]{C_{m^\prime m}(\alpha)}
\newcommand{\Cmmi}[0]{C_{m^\prime m}(\alpha_i)}
\newcommand{\mam}[0]{m\alpha m^\prime}
\newcommand{\blm}[0]{b_{l m^\prime}}
\newcommand{\Nbin}[0]{N_{\text{bin}}}
\newcommand{\slm}[0]{s_{l m}}
\newcommand{\ylm}[0]{Y_{l m}}
\newcommand{\dconv}[0]{d^l_{m^\prime m}(\alpha)}
\newcommand{\dconvi}[0]{d^l_{m^\prime m}(\alpha_i)}
\newcommand{\dconviz}[0]{d^l_{0 m}(\alpha_i)}
\newcommand{\dzero}[0]{d^l_{0 m}(\alpha_i)}
\newcommand{\dconvbar}[0]{d^{\bar{l}}_{m_{\text{b}}m_{\text{sky}}}(\beta)}
\newcommand{\lmax}[0]{l_{\text{max}}}
\newcommand{\mb}[0]{m_{\text{b}}}
\newcommand{\msky}[0]{m_{\text{sky}}}
\newcommand{\mmax}[0]{m_{\text{max}}}
\newcommand{\mbmax}[0]{m_{b\text{max}}}
\newcommand{\mskymax}[0]{m_{\text{sky}\text{max}}}
\newcommand{\Dlmm}[0]{D^l_{mm^\prime}(\psi,\alpha,\phi)}
\newcommand{\eq}[1]{eq.~(\ref{#1})}
\newcommand{\eqs}[2]{eqs.~(\ref{#1,#2})}
\newcommand{\Eq}[1]{Eq.~(\ref{#1})}
\newcommand{\Eqs}[2]{Eqs.~(\ref{#1},\ref{#2})}
\newcommand{\fig}[1]{fig.~\ref{#1}}
\newcommand{\figs}[2]{figs.~\ref{#1},\ref{#2}}
\newcommand{\Fig}[1]{Fig.~\ref{#1}}
\newcommand{\qbb}[0]{Q_{\beta\beta}}
\newcommand{\tl}[0]{\text{L}}
\newcommand{\tr}[0]{\text{R}}
\newcommand{\nc}[0]{N_{\text{c}}}
\newcommand{\mw}[0]{M_{\text{W}}}
\newcommand{\mz}[0]{M_{\text{Z}}}
\newcommand{\mr}[0]{M_{\text{R}}}
\newcommand{\md}[0]{m_{\text{D}}}
\newcommand{\mn}[0]{m_\nu}
\newcommand{\lh}[0]{\Lambda_{\text{H}}}
\newcommand{\aif}[0]{a^{ff^\prime}_{i;ll^\prime}}
\newcommand{\gs}[0]{\Gamma_{\text{S}}=1}
\newcommand{\gps}[0]{\Gamma_{\text{PS}}=\gamma^5}
\newcommand{\lr}[0]{\Lambda_{\text{R}}}
\newcommand{\wrt}[0]{W_{\text{R}}}
\newcommand{\wl}[0]{W_{\text{L}}}
\newcommand{\ls}[0]{\Lambda_{\text{S}}}
\newcommand{\gf}[0]{G_{\text{F}}}
\newcommand{\mm}[0]{M_{\text{M}}}
\newcommand{\sst}[1]{{\scriptscriptstyle #1}}
\newcommand{\beq}{\begin{equation}}
\newcommand{\eeq}{\end{equation}}
\newcommand{\beqa}{\begin{eqnarray}}
\newcommand{\eeqa}{\end{eqnarray}}
\newcommand{\dida}[1]{/ \!\!\! #1}
\renewcommand{\Im}{\mbox{\sl{Im}}}
\renewcommand{\Re}{\mbox{\sl{Re}}}
\def\simge{\hspace*{0.2em}\raisebox{0.5ex}{$>$}
     \hspace{-0.8em}\raisebox{-0.3em}{$\sim$}\hspace*{0.2em}}
\def\simle{\hspace*{0.2em}\raisebox{0.5ex}{$<$}
     \hspace{-0.8em}\raisebox{-0.3em}{$\sim$}\hspace*{0.2em}}
\def\dn{{d_n}}
\def\de{{d_e}}
\def\datom{{d_{\sst{A}}}}
\def\grhobar{{{\bar g}_\rho}}
\def\gpibar{{{\bar g}_\pi^{(I) \prime}}}
\def\gpibarz{{{\bar g}_\pi^{(0) \prime}}}
\def\gpibaro{{{\bar g}_\pi^{(1) \prime}}}
\def\gpibart{{{\bar g}_\pi^{(2) \prime}}}
\def\mx{{M_X}}
\def\mrho{{m_\rho}}
\def\qpv{{Q_{\sst{W}}}}
\def\lamtv{{\Lambda_{\sst{TVPC}}}}
\def\lamtvs{{\Lambda_{\sst{TVPC}}^2}}
\def\lamtvc{{\Lambda_{\sst{TVPC}}^3}}

\def\bra#1{{\langle#1\vert}}
\def\ket#1{{\vert#1\rangle}}
\def\coeff#1#2{{\scriptstyle{#1\over #2}}}
\def\undertext#1{{$\underline{\hbox{#1}}$}}
\def\hcal#1{{\hbox{\cal #1}}}
\def\sst#1{{\scriptscriptstyle #1}}
\def\eexp#1{{\hbox{e}^{#1}}}
\def\rbra#1{{\langle #1 \vert\!\vert}}
\def\rket#1{{\vert\!\vert #1\rangle}}

\def\lsim{{ <\atop\sim}}
\def\gsim{{ >\atop\sim}}
\def\nubar{{\bar\nu}}
\def\psibar{{\bar\psi}}
\def\Gmu{{G_\mu}}
\def\alr{{A_\sst{LR}}}
\def\wpv{{W^\sst{PV}}}
\def\evec{{\vec e}}
\def\notq{{\not\! q}}
\def\notl{{\not\! \ell}}
\def\notk{{\not\! k}}
\def\notp{{\not\! p}}
\def\notpp{{\not\! p'}}
\def\notder{{\not\! \partial}}
\def\notcder{{\not\!\! D}}
\def\notA{{\not\!\! A}}
\def\notv{{\not\!\! v}}
\def\Jem{{J_\mu^{em}}}
\def\Jana{{J_{\mu 5}^{anapole}}}
\def\nue{{\nu_e}}
\def\mns{{m^2_{\sst{N}}}}
\def\mes{{m^2_e}}
\def\mq{{m_q}}
\def\mqs{{m_q^2}}
\def\mw{{M_{\sst{W}}}}
\def\mz{{M_{\sst{Z}}}}
\def\mzs{{M^2_{\sst{Z}}}}
\def\ubar{{\bar u}}
\def\dbar{{\bar d}}
\def\sbar{{\bar s}}
\def\qbar{{\bar q}}
\def\sstw{{\sin^2\theta_{\sst{W}}}}
\def\gv{{g_{\sst{V}}}}
\def\ga{{g_{\sst{A}}}}
\def\pv{{\vec p}}
\def\pvs{{{\vec p}^{\>2}}}
\def\ppv{{{\vec p}^{\>\prime}}}
\def\ppvs{{{\vec p}^{\>\prime\>2}}}
\def\qv{{\vec q}}
\def\qvs{{{\vec q}^{\>2}}}
\def\xv{{\vec x}}
\def\xpv{{{\vec x}^{\>\prime}}}
\def\yv{{\vec y}}
\def\tauv{{\vec\tau}}
\def\sigv{{\vec\sigma}}

\def\sst#1{{\scriptscriptstyle #1}}
\def\gpnn{{g_{\sst{NN}\pi}}}
\def\grnn{{g_{\sst{NN}\rho}}}
\def\gnnm{{g_{\sst{NNM}}}}
\def\hnnm{{h_{\sst{NNM}}}}
\def\xivz{{\xi_\sst{V}^{(0)}}}
\def\xivt{{\xi_\sst{V}^{(3)}}}
\def\xive{{\xi_\sst{V}^{(8)}}}
\def\xiaz{{\xi_\sst{A}^{(0)}}}
\def\xiat{{\xi_\sst{A}^{(3)}}}
\def\xiae{{\xi_\sst{A}^{(8)}}}
\def\xivtez{{\xi_\sst{V}^{T=0}}}
\def\xivteo{{\xi_\sst{V}^{T=1}}}
\def\xiatez{{\xi_\sst{A}^{T=0}}}
\def\xiateo{{\xi_\sst{A}^{T=1}}}
\def\xiva{{\xi_\sst{V,A}}}
\def\rvz{{R_{\sst{V}}^{(0)}}}
\def\rvt{{R_{\sst{V}}^{(3)}}}
\def\rve{{R_{\sst{V}}^{(8)}}}
\def\raz{{R_{\sst{A}}^{(0)}}}
\def\rat{{R_{\sst{A}}^{(3)}}}
\def\rae{{R_{\sst{A}}^{(8)}}}
\def\rvtez{{R_{\sst{V}}^{T=0}}}
\def\rvteo{{R_{\sst{V}}^{T=1}}}
\def\ratez{{R_{\sst{A}}^{T=0}}}
\def\rateo{{R_{\sst{A}}^{T=1}}}
\def\mro{{m_\rho}}
\def\mks{{m_{\sst{K}}^2}}
\def\mpi{{m_\pi}}
\def\mpis{{m_\pi^2}}
\def\mom{{m_\omega}}
\def\mphi{{m_\phi}}
\def\Qhat{{\hat Q}}
\def\FOS{{F_1^{(s)}}}
\def\FTS{{F_2^{(s)}}}
\def\GAS{{G_{\sst{A}}^{(s)}}}
\def\GES{{G_{\sst{E}}^{(s)}}}
\def\GMS{{G_{\sst{M}}^{(s)}}}
\def\GATEZ{{G_{\sst{A}}^{\sst{T}=0}}}
\def\GATEO{{G_{\sst{A}}^{\sst{T}=1}}}
\def\mdax{{M_{\sst{A}}}}
\def\mustr{{\mu_s}}
\def\rsstr{{r^2_s}}
\def\rhostr{{\rho_s}}
\def\GEG{{G_{\sst{E}}^\gamma}}
\def\GEZ{{G_{\sst{E}}^\sst{Z}}}
\def\GMG{{G_{\sst{M}}^\gamma}}
\def\GMZ{{G_{\sst{M}}^\sst{Z}}}
\def\GEn{{G_{\sst{E}}^n}}
\def\GEp{{G_{\sst{E}}^p}}
\def\GMn{{G_{\sst{M}}^n}}
\def\GMp{{G_{\sst{M}}^p}}
\def\GAp{{G_{\sst{A}}^p}}
\def\GAn{{G_{\sst{A}}^n}}
\def\GA{{G_{\sst{A}}}}
\def\GETEZ{{G_{\sst{E}}^{\sst{T}=0}}}
\def\GETEO{{G_{\sst{E}}^{\sst{T}=1}}}
\def\GMTEZ{{G_{\sst{M}}^{\sst{T}=0}}}
\def\GMTEO{{G_{\sst{M}}^{\sst{T}=1}}}
\def\lamd{{\lambda_{\sst{D}}^\sst{V}}}
\def\lamn{{\lambda_n}}
\def\lams{{\lambda_{\sst{E}}^{(s)}}}
\def\bvz{{\beta_{\sst{V}}^0}}
\def\bvo{{\beta_{\sst{V}}^1}}
\def\Gdip{{G_{\sst{D}}^\sst{V}}}
\def\GdipA{{G_{\sst{D}}^\sst{A}}}
\def\fks{{F_{\sst{K}}^{(s)}}}
\def\FIS{{F_i^{(s)}}}
\def\fpi{{F_\pi}}
\def\fk{{F_{\sst{K}}}}
\def\RAp{{R_{\sst{A}}^p}}
\def\RAn{{R_{\sst{A}}^n}}
\def\RVp{{R_{\sst{V}}^p}}
\def\RVn{{R_{\sst{V}}^n}}
\def\rva{{R_{\sst{V,A}}}}
\def\xbb{{x_B}}
\def\mlq{{M_{\sst{LQ}}}}
\def\mlqs{{M_{\sst{LQ}}^2}}
\def\lscal{{\lambda_{\sst{S}}}}
\def\lvect{{\lambda_{\sst{V}}}}
\def\PR#1{{{\em   Phys. Rev.} {\bf #1} }}
\def\PRC#1{{{\em   Phys. Rev.} {\bf C#1} }}
\def\PRD#1{{{\em   Phys. Rev.} {\bf D#1} }}
\def\PRL#1{{{\em   Phys. Rev. Lett.} {\bf #1} }}
\def\NPA#1{{{\em   Nucl. Phys.} {\bf A#1} }}
\def\NPB#1{{{\em   Nucl. Phys.} {\bf B#1} }}
\def\AoP#1{{{\em   Ann. of Phys.} {\bf #1} }}
\def\PRp#1{{{\em   Phys. Reports} {\bf #1} }}
\def\PLB#1{{{\em   Phys. Lett.} {\bf B#1} }}
\def\ZPA#1{{{\em   Z. f\"ur Phys.} {\bf A#1} }}
\def\ZPC#1{{{\em   Z. f\"ur Phys.} {\bf C#1} }}
\def\etal{{{\em   et al.}}}
\def\delalr{{{delta\alr\over\alr}}}
\def\pbar{{\bar{p}}}
\def\lamchi{{\Lambda_\chi}}
\def\qw0{{Q_{\sst{W}}^0}}
\def\qwp{{Q_{\sst{W}}^P}}
\def\qwn{{Q_{\sst{W}}^N}}
\def\qwe{{Q_{\sst{W}}^e}}
\def\qem{{Q_{\sst{EM}}}}
\def\gae{{g_{\sst{A}}^e}}
\def\gve{{g_{\sst{V}}^e}}
\def\gvf{{g_{\sst{V}}^f}}
\def\gaf{{g_{\sst{A}}^f}}
\def\gvu{{g_{\sst{V}}^u}}
\def\gau{{g_{\sst{A}}^u}}
\def\gvd{{g_{\sst{V}}^d}}
\def\gad{{g_{\sst{A}}^d}}
\def\gvftil{{\tilde g_{\sst{V}}^f}}
\def\gaftil{{\tilde g_{\sst{A}}^f}}
\def\gvetil{{\tilde g_{\sst{V}}^e}}
\def\gaetil{{\tilde g_{\sst{A}}^e}}
\def\gvqtil{{\tilde g_{\sst{V}}^e}}
\def\gaqtil{{\tilde g_{\sst{A}}^e}}
\def\gvutil{{\tilde g_{\sst{V}}^e}}
\def\gautil{{\tilde g_{\sst{A}}^e}}
\def\gvdtil{{\tilde g_{\sst{V}}^e}}
\def\gadtil{{\tilde g_{\sst{A}}^e}}
\def\delp{{\delta_P}}
\def\delzp{{\delta_{00}}}
\def\deld{{\delta_\Delta}}
\def\dele{{\delta_e}}
\def\lnew{{{\cal L}_{\sst{NEW}}}}
\def\osffp{{{\cal O}_{7a}^{ff'}}}
\def\oszg{{{\cal O}_{7c}^{Z\gamma}}}
\def\osgg{{{\cal O}_{7b}^{g\gamma}}}


\def\slash#1{#1\!\!\!{/}}
\def\beq{\begin{eqnarray}}
\def\eeq{\end{eqnarray}}
\def\bea{\begin{eqnarray*}}
\def\eea{\end{eqnarray*}}
\def\NCA{\em Nuovo~Cimento}
\def\IJMP{\em Intl.~J.~Mod.~Phys.}
\def\NP{\em Nucl.~Phys.}
\def\PLB{{\em Phys.~Lett.}~B}
\def\JETPLett{{\em JETP Lett.}}
\def\PRL{\em Phys.~Rev.~Lett.}
\def\MPL{\em Mod.~Phys.~Lett.}
\def\PRD{{\em Phys.~Rev.}~D}
\def\PR{\em Phys.~Rev.}
\def\PRP{\em Phys.~Rep.}
\def\ZPC{{\em Z.~Phys.}~C}
\def\PTP{{\em Prog.~Theor.~Phys.}}
\def\Baryon{{\rm B}}
\def\Lepton{{\rm L}}
\def\sbar{\overline}
\def\stilde{\widetilde}
\def\st{\scriptstyle}
\def\sst{\scriptscriptstyle}
\def\vac{|0\rangle}
\def\argh{{{\rm arg}}}
\def\G{\stilde G}
\def\Wmess{W_{\rm mess}}
\def\NI{\stilde N_1}
\def\antivac{\langle 0|}
\def\infinity{\infty}
\def\mco{\multicolumn}
\def\epp{\epsilon^{\prime}}
\def\psibar{\overline\psi}
\def\nmess{N_5}
\def\chibar{\overline\chi}
\def\lagr{{\cal L}}
\def\drbar{\overline{\rm DR}}
\def\msbar{\overline{\rm MS}}
\def\conj{{{\rm c.c.}}}
\def\Et{{\slashchar{E}_T}}
\def\Etot{{\slashchar{E}}}
\def\mZ{m_Z}
\def\MPlanck{M_{\rm P}}
\def\mW{m_W}
\def\cbeta{c_{\beta}}
\def\sbeta{s_{\beta}}
\def\cW{c_{W}}
\def\sW{s_{W}}
\def\deltaeps{\delta}
\def\sigmabar{\overline\sigma}
\def\epsilonbar{\overline\epsilon}
\def\vep{\varepsilon}
\def\ra{\rightarrow}
\def\half{{1\over 2}}
\def\ko{K^0}
\def\be{\beq}
\def\ee{\eeq}
\def\bea{\begin{eqnarray}}
\def\eea{\end{eqnarray}}
\def\alr{A_{\sst{LR}}}

\def\centeron#1#2{{\setbox0=\hbox{#1}\setbox1=\hbox{#2}\ifdim
\wd1>\wd0\kern.5\wd1\kern-.5\wd0\fi
\copy0\kern-.5\wd0\kern-.5\wd1\copy1\ifdim\wd0>\wd1
\kern.5\wd0\kern-.5\wd1\fi}}
\def\ltap{\;\centeron{\raise.35ex\hbox{$<$}}{\lower.65ex\hbox{$\sim$}}\;}
\def\gtap{\;\centeron{\raise.35ex\hbox{$>$}}{\lower.65ex\hbox{$\sim$}}\;}
\def\gsim{\mathrel{\gtap}}
\def\lsim{\mathrel{\ltap}}
\def\slashchar#1{\setbox0=\hbox{$#1$}           
   \dimen0=\wd0                                 
   \setbox1=\hbox{/} \dimen1=\wd1               
   \ifdim\dimen0>\dimen1                        
      \rlap{\hbox to \dimen0{\hfil/\hfil}}      
      #1                                        
   \else                                        
      \rlap{\hbox to \dimen1{\hfil$#1$\hfil}}   
      /                                         
   \fi}                                        %

\setcounter{tocdepth}{2}







\title{Dense Dark Matter Hairs Spreading Out from Earth, Jupiter and Other Compact Bodies} 

\author{G.~Pr{\'e}zeau}
\address {Jet Propulsion Laboratory, California Institute of Technology, 4800 Oak Grove Dr., Pasadena, CA 91109, USA}




\begin{abstract}
It is shown that compact bodies project out strands of concentrated dark matter filaments henceforth simply called hairs.  These hairs are a consequence of the fine-grained stream structure of dark matter halos, and as such constitute a new physical prediction of   $\Lambda$CDM.  Using both an analytical model of planetary density and  numerical simulations utilizing the  {\it Fast Accurate Integrand Renormalization } (FAIR) algorithm (a fast geodesics calculator described below) with realistic planetary density inputs, dark matter streams moving through  a compact body are shown to produce hugely magnified dark matter densities  along the stream velocity axis going through the center of the body.  Typical hair density enhancements are $10^7$ for Earth and $10^8$ for Jupiter.  The largest enhancements occur for particles streaming through the core of the body that mostly focus at a single point called the root of the hair.  For the Earth, the root is located at about $10^6$~km from the planetary center with a density enhancement of around $10^9$  while for a gas giant like Jupiter, the root is located at around $10^{5}$~km with a enhancement of around $10^{11}$.  Beyond the root, the hair density precisely reflects the density layers of the body providing a direct probe of planetary interiors.
\end{abstract}

\section{Introduction}

As the evidence of dark matter's existence has become overwhelming\footnote{In particular, the accumulation of evidence for dark matter's existence spans generations~\citep{Zwicky:1933gu,1973ApJ...186..467O,1974ApJ...193L...1O,1980ApJ...241..552F} with more recent and striking observations such as the bullet cluster~\citep{Clowe:2006eq}.  For a discussion of the challenges facing the Modified Newtonian Dynamics (MOND) alternative to dark matter, see~\cite{Dodelson:2011qv}.  For a look at the constraints on warm dark matter, see \cite{Viel:2013apy} and \cite{Anderhalden:2012jc}.   For reviews of the evidence of dark matter, see \cite{2010pdmo.book.....B} and \cite{Massey:2010hh}; see also the latest Planck results for the cosmic microwave background evidence for cold dark matter~~\cite{Ade:2015xua}.},  understanding its nature and interactions through the observation of both direct (from Earth-bound experiments) and indirect (excess $\gamma$-ray and positron measurements) dark matter detection has become an increasingly vigorous field of inquiry.  In addition, upper-limits on dark matter interactions are now extracted from the analysis of large samples of colliding clusters~\citep{Harvey:2015hha}, and  constraints on both dark matter stream velocity dispersions and  halo formation  can be deduced from large structures like the "Field of Streams"~\citep{Belokurov:2006ms}.  In spite of all these extraordinary advances, little is known about the dark matter particle itself (whether it is a boson, Dirac or Majorana fermion), its local density and velocity distribution in the solar system, or its interactions.  This missing information  may  remain hidden until a direct detection is made and their interactions can be studied consistently over time; a longitudinal study could be performed thanks to dark matter production at an accelerator or  because a naturally occurring concentrated dark matter source is discovered locally that can be reliably accessed with instruments. Highly concentrated dark matter hairs would be such a source and are in fact a prediction of $\Lambda$CDM, a model that fits very well with the most recent Planck CMB analysis and many other astrophysical data\footnote{See section 5 of Ref.~\cite{Ade:2015xua} and references therein.}.

For thermally produced Cold Dark Matter  (CDM) of the type allowed by $\Lambda$CDM, the CDM primordial thermal velocity dispersion is expected to be greatly suppressed as the universe expands and the CDM collisionless gas cools.  In particular, for a WIMP with mass 100~GeV that decoupled at 10~MeV, the velocity dispersion is about 3$\times10^{-4}$~m/s while the velocity dispersion for a 10~$\mu$eV axion is $10^{-8}$~m/s~\citep{Sikivie:1999jv}.  As the non-linear effects of gravity become more prominent and the halos grow, a coarse-grained velocity dispersion of the CDM will appear as they orbit the galaxy~\citep{Sikivie:1995dp}; the coarseness of this dispersion will gradually smooth out as the number of  orbits increases.  In the ``Field of Streams"~\citep{Belokurov:2006ms}, the {\it effective} velocity dispersion is 10~km/s providing an experimental upper-limit on the velocity dispersion of the fine-grained dark matter streams, each of which should have a tiny primordial velocity dispersion.  The ``Field of Streams" is an example of a {\it tidal} dark matter stream built-up from a huge number of fine-grained dark matter streams and distinct from them. A phase-space perspective sheds additional light on the processes affecting the CDM under the influence of gravity.

At the last scattering surface, the CDM occupy a 3-dimensional sheet in the 6-dimensional phase space since they have  tiny velocity dispersions.  The process of galactic halo formation cannot tear this hypersurface filled with the collisionless gas of CDM because of the generalized Liouville's theorem that accounts for the time evolution of the metric.  Under the influence of gravity, a particular phase space volume of the hypersurface is stretched and folded with each orbit of the CDM creating layers of fine-grained dark matter streams, each with a vanishingly small velocity dispersion.  These stretches and folds also produced  caustics~\citep{Tremaine:1998nk}, regions with very high CDM densities that are inversely proportional to the square root of the velocity dispersion, potentially providing large boosts to CDM annihilation into photons.  Identifying such boosts is important because it would help  explain the positron excess seen by PAMELA for energies above $10$~GeV~\citep{Adriani:2013uda}, if such excess stems from WIMP annihilation.  Indeed, supersymmetric radiative corrections to the neutralino annihilation positron branching ratio can explain the rise in positron excess for rising energies provided features (like caustics) in the halo structure  boost the dark matter density~\citep{Bergstrom:2008gr}.

In order to understand how caustics could provide such a boost, Vogelsberger and White~\citep{Vogelsberger:2010gd} (henceforth V\&W) integrated the geodesic deviation equation~\citep{Vogelsberger:2007ny} in tandem with the equations of motion of $N$ simulation particles to analyze the fine-grained stream structure of the halos starting from general $\Lambda$CDM initial conditions.  They were able to count all the caustics encountered by a simulation particle along its geodesic and calculate the CDM annihilation rate enhancements; they found  that  caustics provide at best a boost of 0.1\%.  The V\&W simulations also produced a probability distribution that a fine-grained stream containing a particular fraction of the average local density would pass through a detector on Earth.  For example, they found that there was a 20\% chance of a local CDM detection where a single stream accounted for  1\% of the total signal.  These results led V\&W to conclude that the CDM velocity distribution was locally smooth and that the fine-grained halo structure would be too difficult to detect on Earth unless the CDM are axions.

On the other hand,  space-based detectors could discern the local fine-grained structure because the Earth  acts as a gravitational lens that separates  dark matter streams\footnote{From hereon, dark matter stream refers exclusively to the primordial dark matter streams with tiny velocity dispersions, and not to tidal dark matter streams unless explicitly noted.} with different, sharply-peaked, velocities.  In this paper, a Schwarzschild metric is used to show that weakly interacting particles streaming at  220~km/s (the approximate orbital velocity of the solar system around the galactic center~\citep{1996AstL...22..455K}) through a compact body with a small impact parameter (i.e., near the core)  will experience huge CDM density enhancements at the root (the nearest focal point composed of particles streaming through the core of the body) of the order of $10^9$ for the Earth and $10^{11}$ for Jupiter.  For the Sun, the entire hair from root to tip (the focal point of particles streaming at the surface of a body)\footnote{It will be seen below that the flux of massive particles is also magnified for particles external to a body suggesting that the hair length may be infinite.  However, the focal point distance increases too fast for external solutions undermining their experimental usefulness.} is inside the solar radius for median CDM velocities of 220~km/s so that any direct detection near the Sun would be suppressed by the low probability of finding a high velocity dark matter stream, not to mention the challenges of close proximity to the Sun.  Dense dark matter hairs provide an entirely unique opportunity to study dark matter interactions and local stream properties.  These CDM density enhancements will boost their direct detection rate for a space-bound experiment without invoking any physics beyond the standard model of particle physics, $\Lambda$CDM and general relativity and may be the only way to obtain empirical data on local dark matter streams which would further our understanding of halo structure and formation.

This paper is organized as follows:  In  section 2, known results for exterior geodesic solutions near Earth of both massless and massive particles are briefly reviewed for completeness and a description of the flux magnification for external solutions is given.  Section 3 derives the analytic, constant density interior solutions of the geodesic equation (GE) and the density enhancement for massive particles is shown to be huge; the massive particle solution is expanded in powers of the impact parameter $b$, and the focal point is found to be independent of $b$ to leading order with corrections of order $(b/\re)^2$ where $\re$ is the radius of the Earth (but could be the radius of any compact body). It is also shown that  the non-zero but tiny velocity dispersion of the CDM has a negligible impact on the density enhancements of the dark matter hairs for standard mass assignments of the CDM.  In section~4, the {\it Fast Accurate Integrand Renormalization} (FAIR) algorithm  for solving the GE for realistic radial densities is described and compared with the analytic constant density result.  Because FAIR is an algorithm that modifies the short scale radial dependence of the density profile, it can be used to demonstrate the robustness of the density enhancement results.   In section 5, the numerical solutions of the GE for a model Earth with a radial density given by the Preliminary Reference Earth Model (PREM)~\citep{1981PEPI...25..297D} are calculated using FAIR, and the hair focal points and density enhancements are plotted as a function of the impact parameter; the same plots are generated for Jupiter with a radial density given by the model J11-4a from Ref.~\cite{Nettelmann:2011kx}.  In the last section, the implications of our results are discussed, including the benefits of  exploring a hair (which includes a powerful new tool to explore the interiors of compact bodies) and the conditions constraining a space mission dedicated to finding such a hair; an equation relating the different parameters of such a mission is also derived.

\section{Exterior Solutions and Density Enhancement for Earth}

As a weakly interacting particle streams above the surface of the Earth, its trajectory curves towards a point on the axis going through the center of mass of the Earth and parallel to the incident velocity at $t\rightarrow-\infinity$.  The location of this point and the corresponding magnification of flux intensity  can be extracted from the known geodesic solutions for massless and massive particles in a Schwarzschild metric. The geodesic equation for a massive particle is  given by
\beq
c^2 &=& \left(1-\frac{\rs}{r}\right)  c^2  \left(\frac{\text{d}t}{\text{d}\tau}\right)^2 - \left(   1-\frac{\rs}{r} \right)^{-1} \left(\frac{\text{d}r}{\text{d}\tau}\right)^2
  \nonumber \\
& & ~~~~~~~~~~~~~~~~~~~~~~~ -  r^2\sin^2\theta \left(\frac{\text{d}\phi}{\text{d}\tau}\right)^2  - r^2 \left(\frac{\text{d}\theta}{\text{d}\tau}\right)^2 ~,
\eeq
while the equation for massless particles is 
\beq
\left(1-\frac{\rs}{r}\right)  c^2  \left(\frac{\text{d}t}{\text{d}\tau}\right)^2 &=& \left(   1-\frac{\rs}{r} \right)^{-1} \left(\frac{\text{d}r}{\text{d}\tau}\right)^2
  \nonumber \\
& & ~~~~~~~~~~ +  r^2\sin^2\theta \left(\frac{\text{d}\phi}{\text{d}\tau}\right)^2  + r^2 \left(\frac{\text{d}\theta}{\text{d}\tau}\right)^2 ~,
\eeq
where $\rs=2G\me/c^2$ is the Schwarzschild radius, $\me$ is the Earth's mass and $G$ the gravitational constant.  The $\tau$ derivative of $\theta$  is zero thanks to angular momentum conservation with $\sin\theta=1$.

The exterior solution for a massless particle with impact parameter $b>\re$ can be obtained by integrating
\beq\label{extmasseq}
\frac{\text{d}\phi}{\text{d}u} = -b\left( 1 - b^2u^2 +\rs b^2u^3    \right)^{-\frac{1}{2}}
\eeq
giving the usual solution
\beq
\phi&=&\phi^\prime + \text{acos}(bu) + \frac{\rs}{b}    ~,  \\
\phi^\prime &=& \frac{\pi}{2} - \frac{\rs}{b}
\eeq
where $u\equiv 1/r$; this corresponds to a deflection angle
\beq
\delta\phi=2\frac{\rs}{b}.
\eeq
Since $\rs=9\times 10^{-3}$m for the Earth and  $b>\re=6.371\times 10^6$ for geodesics external to the Earth, the deflection of massless non-interacting particles is effectively negligible with focal points
\beq
F(b)=\frac{b^2}{2\rs} > 2\times 10^{15}~\text{m}
\eeq
which are completely unreachable.  The exterior solutions for massless particles with small impact parameters ($b\le\re$) exterior to the Earth will be needed below for the initial angle value of the interior solutions of massless particles.  They can be obtained by expanding the denominator in \Eq{extmasseq} and integrating yielding
\beq
\phi = \pi - \frac{\rs}{b}\left( 1  - \sqrt{1-b^2 u^2} \right)  -  \text{arcsin}(bu)~.
\eeq
For massive particles neglecting the insignificant terms proportional to $\rs/r$ the exterior solution is the usual orbit equation
\beq
\frac{1}{r}&=&\frac{1}{\xi b^2}\left[   1 + \sqrt{ 1 + \xi^2 b^2}\cos(\phi-\phi^\prime_{\text{E}}) \right] \\
\phi^\prime_{\text{E}} &=&\text{acos}\left ( 1 + \xi^2 b^2  \right)^{-\frac{1}{2}} ~, \\
\xi &\equiv&   \frac{v^2}{G\me} ~.
\eeq
for particle velocity $v$.  If $v=220$~km/h,
\beq
 & &\xi^2 b^2 > \xi^2 \re^2 =6\times 10^5~, \\
 & &\phi^\prime_{\text{E}} \cong  \text{acos}\left ( \frac{G\me}{b v^2} \right)\cong\frac{\pi}{2}-\frac{1}{b\xi}~.
\eeq
\begin{figure}
\resizebox{9cm}{!}{\includegraphics*{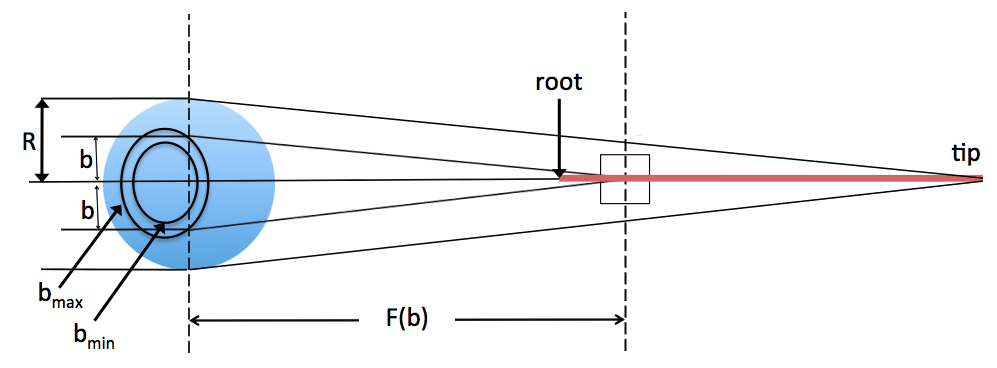}}
\caption{Simplified drawing (not to scale) of DM particles, belonging to a single stream,  passing through a body.  The hair is depicted in red, from the root (where the particles streaming near the core focus) to the tip (where the particles grazing the surface of a body of  radius $R$ focus). Figure~\ref{detPic} shows an inset of the square above, where dark matter particles with impact parameter  $b_{\text{min}}>b>b_{\text{max}}$ focus near $F(b)$.  All the particles belonging to the same stream, and passing through the ring with perimeters located at $b_{\text{min}}$ and $b_{\text{max}}$, will  focus on a circular detector with radius $R_{\text{D}}$.}\label{bigPic}
\end{figure}
\begin{figure}
\resizebox{9cm}{!}{\includegraphics*{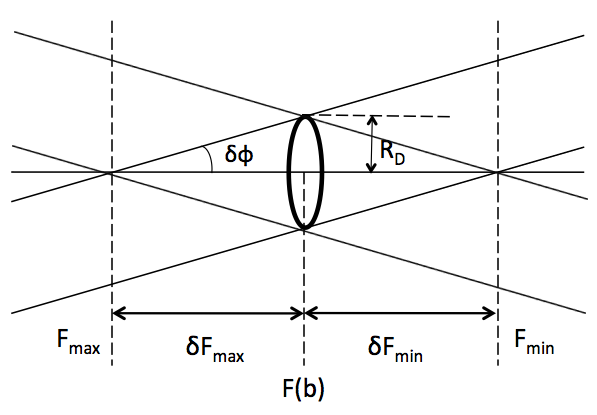}}
\caption{All particles from a dark matter stream with impact parameter between $b_\text{max}$ and $b_\text{min}$ will cross the axis between $ F(b_{\text{min}})$ and $ F(b_{\text{max}})$ and  stream through (or hit) the circular detector of radius $R_{\text{D}}$.}\label{detPic}
\end{figure}
To calculate the density enhancements by a planetary body of  weakly interacting particles belonging to a dark matter stream\footnote{All dark matter streams in this paper are defined  by their  velocity, $\vec{v}$, and density ratio, $\rho_{\text{s}}/\langle\rho\rangle$, where $\rho_\text{s}$ is the stream density and  $\langle\rho\rangle$ is the local average density.},  a circular-disk detector with radius $R_{\text{D}}$  is taken to be located on the axis going through the center of the planet and parallel to $\vec{v}$ (see Figs.~\ref{bigPic} and \ref{detPic}). In the Schwarzschild metric, every particle belonging to that dark matter stream  will eventually cross that axis.  The density enhancement $M$ at a detector situated at point $F(\bar{b})$ along the axis is given by the ratio of the annulus surface to the surface of the detector
\beq
M[F(b)] =  \frac{b^2_\text{max}-b^2_\text{min}}{\rd^2}=\frac{\delta b^2}{\rd^2}~.
\eeq
Integrating over all possible contributions to a detector located at a point $\bar{b}$ yields
\beq\label{mageq}
M(\bar{b}) &=& \int_0^{\bar{b}} \frac{\text{d}b^2}{\rd^2}\theta\left[ F(b) + \frac{\rd}{\delta\phi(b)} - F(\bar{b}) \right] + \nonumber \\
& & ~~~~~~~~~~~~~~ + \int_{\bar{b}}^{b_{\text{max}}} \frac{\text{d}b^2}{\rd^2}\theta\left[  F(\bar{b}) + \frac{\rd}{\delta\phi(b)} - F(b) \right]  \nonumber  \\
&=& \int_0^{\bar{b}} \frac{\text{d}b^2}{\rd^2}\theta\left[ \frac{F(b)- F(\bar{b})}{\rd}  \left. \frac{\text{d} b }{\text{d} F   }\right|_b + 1   \right] + \nonumber \\
& & ~~~~~~~~~~~~~~ + \int_{\bar{b}}^{b_{\text{max}}} \frac{\text{d}b^2}{\rd^2}\theta\left[  \frac{F(\bar{b})- F(b)}{\rd}  \left. \frac{\text{d} b }{\text{d} F   }\right|_b + 1  \right] 
\eeq
where the step function $\theta(x) = {0,1}~\text{for}~ x<0 ~\text{and}~x\ge 0$ respectively.   Particle moving with speed $v$ with $\bar{b}>\re$ experience a flux enhancement:
\beq
& &\delta\phi=\frac{2}{\bar{b}\xi}~,  \\
& &F(\bar{b})=\frac{\bar{b}^2\xi}{2}>2.5\times10^9~\text{m}~.\\
& &M(\bar{b})=\frac{(\bar{b}+\delta b)^2-(\bar{b}-\delta{b})^2}{\rd^2} = 2\frac{\bar{b}}{\rd}~.
\eeq
Although this is a large enhancement for $b>\re$, the focal point distance increases quadratically with the impact parameter, making it very difficult to benefit from the linearly increasingly magnified flux of streaming massive particles.  In addition, the interior solutions of the geodesic equation of massive particles streaming through the core will be found to be orders of magnitude larger, with focal points located within a million kilometers of the Earth and less for Jupiter.

\section{Interior Solutions: Constant Density Earth}

The interior geodesics of massless particles in a general Schwarzschild metric are solutions of
\beq
e^{-2\Phi/c^2} - \frac{b^2}{r^2} = \left[1-\frac{r_{\text{s}}}{r}\frac{M(r)}{M_{\text{E}}}\right]^{-1}\left(\frac{ \text{d}r }{ \text{d}\phi } \right)^2 \frac{b^2}{r^4}
\eeq
while the massive particles are solutions of
\beq\label{massgeoeq}
\left(  1 + \frac{v^2}{c^2}  \right) e^{-2\Phi/c^2} \!- 1\! - \frac{v^2b^2}{c^2 r^2} = \left[1-\frac{\rs}{r}\frac{M(r)}{M_{\text{E}}}\right]^{-1} \!  \left(\frac{ \text{d}r }{ \text{d}\phi } \right)^2 \! \frac{v^2b^2}{c^2r^4}~.
\eeq
where $\Phi$ is the interior potential of a body.  For compact bodies with constant density, $e^{-\Phi/c^2}$ has an analytical form \citep{wald1984general}
\beq
& &e^{\Phi/c^2} = \frac{3}{2}\sqrt{ 1 - \frac{\rs}{\re}  }  - \frac{1}{2}\sqrt{  1 - \frac{\rs r^2}{\re^3}   } ~, \\ \label{phiC}
& &\Phi\cong-2\pi G\bar{\rho}\left(\re^2-\frac{r^2}{3}\right) ~, \\ \label{meC}
& &M(r)=\frac{4\pi}{3}\bar{\rho} r^3
\eeq
with $\bar{\rho}$ the average density of the body.
\\
\\
{\bf MASSLESS PARTICLES.} In this analytical case, the  differential equation for massless particles is 
\beq\label{m0}
\frac{\text{d}u}{\text{d}\phi} &=& -\frac{1}{u}\left[ -u^4  + su^2 - \frac{3}{2}\frac{\rs}{b^2\re^3} \right]^{\frac{1}{2}}~ , \\
s&\equiv&\left(  \frac{1}{b^2} +\frac{3}{2}\frac{\rs}{b^2\re} +\frac{\rs}{\re^3}  \right) ~,
\eeq
with a  solution that can be readily written down
\beq
\phi &=& \phi^\prime_{\text{I}} +\frac{1}{2}\text{acos}\left[  \frac{  u^2 - \frac{1}{2}s }{  \sqrt{  \frac{1}{4}s^2  - \frac{3}{2}\frac{\rs}{b^2\re^3}  } } \right] ~, \\ \nonumber 
\phi_{\text{I}}^\prime &=&\pi - \frac{\rs}{b}\left(  1 - \sqrt{1-\frac{b^2}{\re^2}} \right) \\ 
& & - \text{arcsin}\!\left[  \! \frac{b}{\re}\left( \! 1-\frac{\rs}{2\re} \! \right) \! \right] - \frac{1}{2} \text{acos}\! \left[ \! \frac{   2/(\re^2s) -1 }{  \sqrt{  1-6\rs b^2 / \re^3   }  } \!  \right] ~ , \\
\delta\phi&=&2\left(\frac{\pi}{2} - \phi_{\text{I}}^\prime\right)~.
\eeq
Note that as  $b\rightarrow \re$, $\phi_{\text{I}}^\prime\rightarrow\phi^\prime$, and that as $b\rightarrow 0$, $\phi_{\text{I}}^\prime\rightarrow \pi/2$, as should be the case.
\begin{figure}
\resizebox{9cm}{!}{\includegraphics*{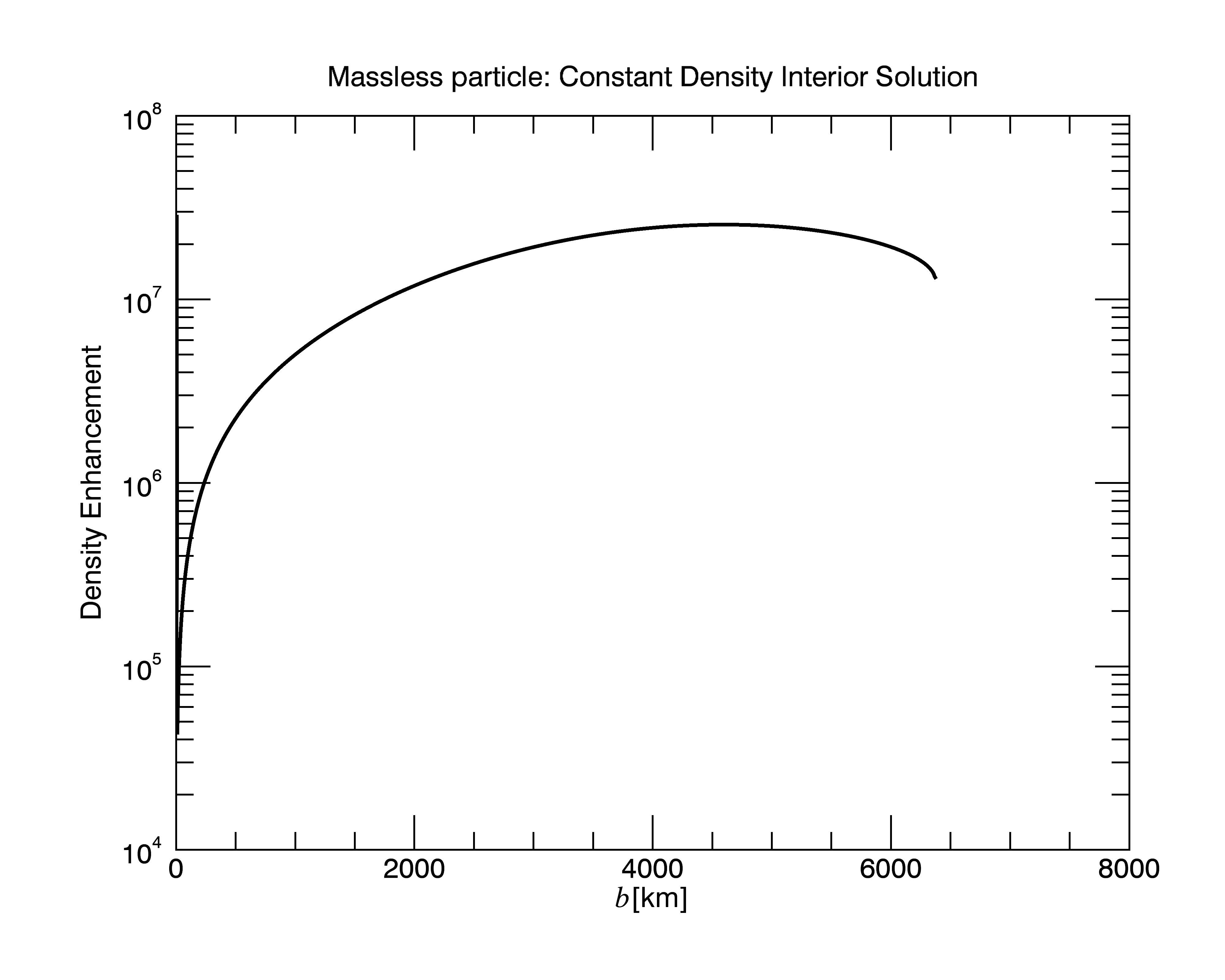}}
\caption{Density enhancement of massless particles streaming through a constant density Earth.  Note that there is a barely visible peak for tiny impact parameter that is more salient in the log-log plot of \Fig{magfpmassless}.}\label{magmassless}
\end{figure}
\begin{figure}
\resizebox{9cm}{!}{\includegraphics*{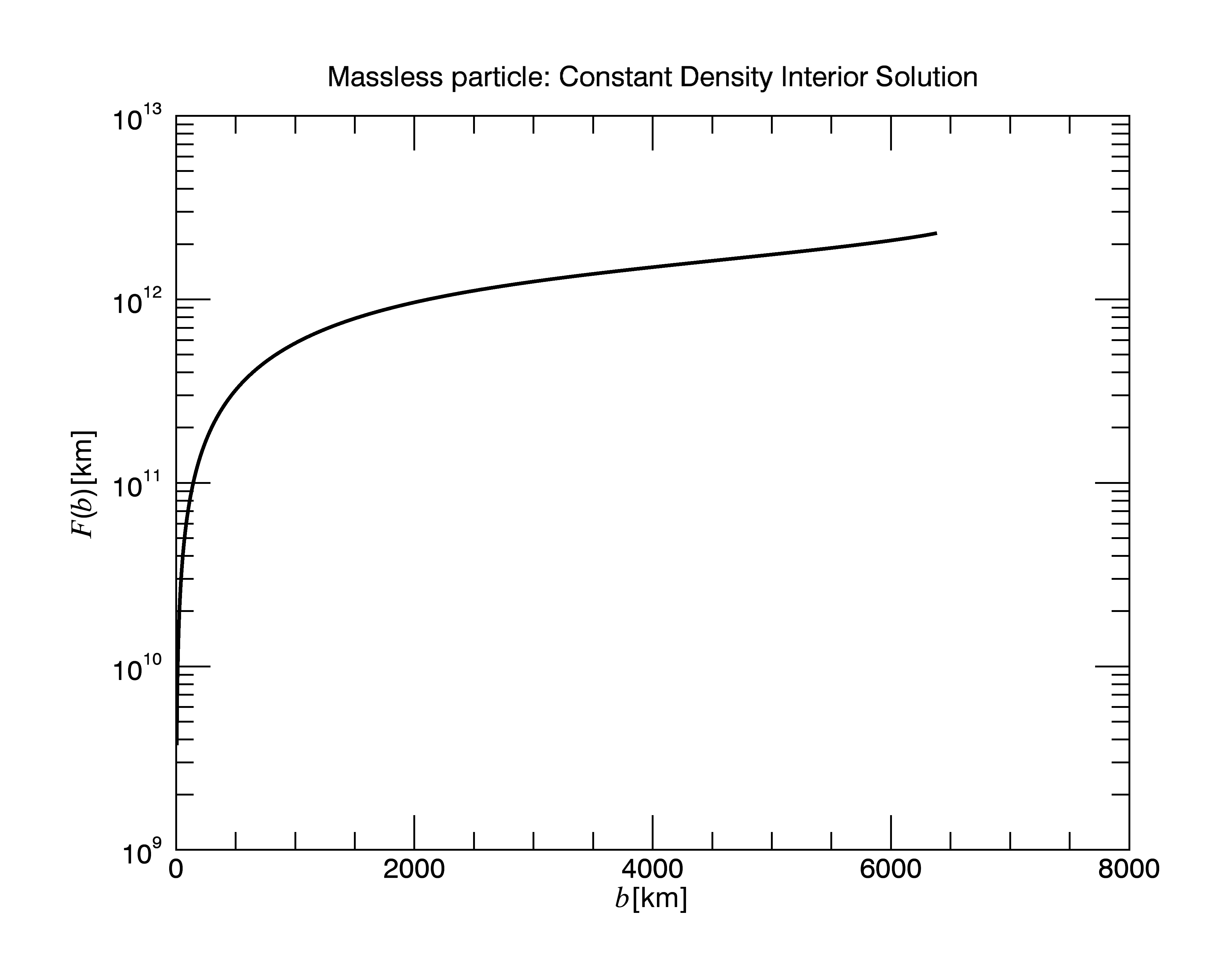}}
\caption{Focal points as a function of impact parameter of massless particles streaming through a constant density Earth.}\label{masslessfp}
\end{figure}
\begin{figure}
\resizebox{9cm}{!}{\includegraphics*{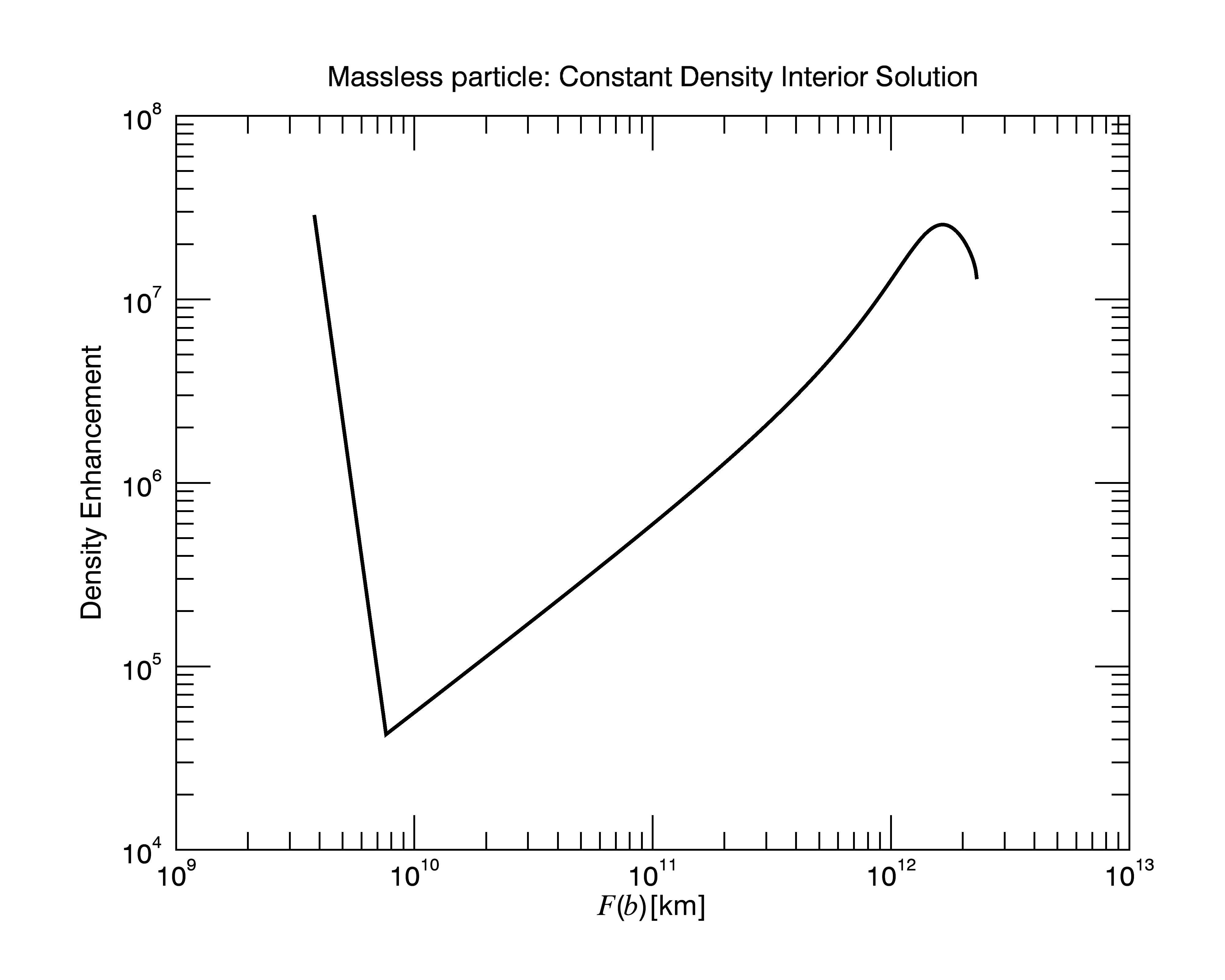}}
\caption{Density enhancement of massless particles streaming through a constant density Earth as a function of focal point distance.}\label{magfpmassless}
\end{figure}
Although the flux of massless particles streaming through a constant density Earth is greatly magnified as seen in \Fig{magmassless},  their focal points  are located far outside the solar system (\Fig{masslessfp})  making explicit the fact that planetary lensing and density enhancement of massless weakly interacting particles cannot be exploited locally for detection purposes\footnote{For massless or high energy neutrinos streaming through the Sun, the focal points are near the orbit of Uranus~\citep{Gerver:1988zs,Escribano:2001ew}}.  In \Fig{magfpmassless}, the log-log plot of the enhancement  as a function of the focal point distance is also given where the enhancement peak for massless particles streaming near the core is now clearly visible (it is pressed against the ordinate in \Fig{magmassless}). In addition, the approximate straight line in the log-log plot between the enhancement and the focal distance shows that any exponential density enhancement is accompanied by exponentially increasing distances.
\\
\\
{\bf MASSIVE PARTICLE.} Using \Eq{massgeoeq}, \Eq{phiC} and \Eq{meC}, the radial equation of motion for a massive particle is given by
\beq\label{constEq}
\frac{\text{d}u}{\text{d}\phi} &=& -\frac{1}{u}\left( -u^4 + \alpha u^2 - \beta \right)^{\frac{1}{2}} ~, \\
\alpha &\equiv& \frac{1}{b^2} +  \frac{3}{2}\frac{\rs}{\re b^2}\left(1+\frac{c^2}{v^2}\right) +  \frac{\rs}{\re^3} ~, \\
\beta &\equiv & \frac{\rs}{2\re^3b^2} \left(3 + \frac{c^2}{v^2} \right) ~.
\eeq
This is almost identical to \Eq{m0} except for the $c^2/v^2$ which now dominates and enhances the $\rs$ terms.  The massive constant density interior solution  is therefore
\beq
\phi &=& \phi^\prime_{\text{M}} +\frac{1}{2}\text{acos}\left[  \frac{  u^2 - \frac{1}{2}\alpha }{  \sqrt{  \frac{1}{4}\alpha^2  - \beta  } } \right] ~,~~~ u^2\ge 1/\re^2 \\
\phi^\prime_{\text{M}} &=& \phi^\prime_{\text{E}}  +   \text{acos}  \left[ \frac{ \xi b^2/\re- 1 }{\sqrt{1+\xi^2 b^2}}   \right]   -  \frac{1}{2} \text{acos}\left[   \frac{  2/(\alpha \re^2) -1   }{  \sqrt{  1- 4\beta/\alpha^2  }  }  \right]      ~.
\eeq
The corresponding density enhancement and focal point are plotted against impact parameter in Figs.~\ref{massiveConstMag} and \ref{massiveConstFP} respectively.  The massive particles streaming near the center of the planet clearly experience the largest magnification.  To understand why, consider the expansion of $\phi^\prime_{\text{M}}$ in powers of $b/\re$
\beq
\phi^\prime_{\text{M}} &\cong& \frac{\pi}{2} - \frac{3}{2\xi \re}   \frac{b}{\re} +  \frac{3b^3}{8\xi\re^4}~, \\
\delta\phi&=& \frac{3}{\xi \re}  \frac{b}{\re} \left( 1-  \frac{1}{4}\frac{b^2}{\re^2}  \right) 
\eeq
The critical thing to note is that the deflection angle $\delta\phi$ is linear in $b$ with corrections suppressed by factors of $(b/\re)^2$.  As such, near the Earth's center,
\beq\label{constfpeq}
F(b) = \frac{\xi\re^2}{3} \left(1 + \frac{ b^2}{4\re^2}  \right) &\cong& 1.6\times10^9~\text{m}  ~,
\eeq
which is consistent with the exact solution in \Fig{massiveConstFP}.  In other words, the focal point is effectively independent of the impact parameter $b$ for particles streaming through the core within a radius of dozens of km's resulting in a huge density enhancement.  This fact is visually clear in \Fig{massiveConstMagFP} where the enhancement is plotted as a function of the focal point where the root of the hair would appear point-like to a space probe.  To calculate the  density enhancement near the core, it is only necessary to solve for $b$
\beq\label{cdmag}
\delta\phi[F(b) - F(\bar{b})]=\rd~.
\eeq
For example, taking $\bar{b}=0$  yields
\beq\label{magb0}
M(0) = \left[ \frac{ (4\rd\re^2)^{\frac{1}{3} } }{\rd} \right]^2=3\times10^{9}~~~(\rd=1~\text{m})~,
\eeq
which is consistent with \Fig{massiveConstMag} and independent of the DM velocity. Velocity independence is important because in a search for hairs, the density enhancement will only depend on the density ratio of the dark matter stream, $\rho_\text{s}/\langle\rho\rangle$.  In addition, the location of the peak density enhancement (which is about twice the value in \Eq{magb0}) is simply
\beq\label{bpeak}
b_\text{peak}=(4\rd\re^2)^{\frac{1}{3}}\cong 55~\text{km for $\rd$=1~m}.
\eeq
Note also that the singularity at $\rd=0$ does not exist in reality because the non-zero velocity dispersion of the dark matter stream is a hard limit on the focus intensity.  For a WIMP dispersion of $3\times10^{-4}$~m/s assumed to be perpendicular to the direction of the WIMP, you get a total perpendicular distance travelled
\beq
\delta \rd \sim d_\perp = \frac{\delta v}{v} F(b) \cong 6~\text{m}~~\text{at}~~b\sim 10~\text{km}
\eeq
for a constant density Earth in the worst case scenario of an entirely perpendicular dispersion; for a $10~\mu$eV axion,  $d_\perp$ is a fraction of a millimeter.  In other words, reducing the detector radius beyond $\delta\rd$ cannot result in a larger density enhancement and \Eq{magb0} can be rewritten
\beq
M(0) = \left[ \frac{ (4\rd\re^2)^{\frac{1}{3} } }{\rd+\delta\rd} \right]^2~.
\eeq
The velocity dispersion at present age for WIMPs and axions is of order~\citep{Sikivie:1999jv}
\beq
\delta v_{\text{a}} &\sim& 10^{-8} \left(   \frac{10^{-5}\text{eV}}{m_{a}}  \right) \text{m/s} \\
\delta v_{\text{W}} &\sim&  3 \times 10^{-3} \sqrt{ \frac{\text{GeV}}{m_{W}} }  \text{m/s} 
\eeq
It is therefore seen that for $m_{\text{a}}> 10^{-9}$~eV and $m_{\text{W}}> 10^{2}$~GeV, the density enhancement observed by a detector with a radius of a few meters will be little affected by the velocity dispersion of the CDM.

The fact that the deflection angle is linear in $b/\re$ with corrections of order $(b/\re)^3$ is not unique to constant density profiles as will  be verified below for the Earth and Jupiter using realistic density profiles from the literature.  We now turn to an algorithm that will permit us to calculate density enhancements for arbitrary density profiles.
\begin{figure}
\resizebox{9cm}{!}{\includegraphics*{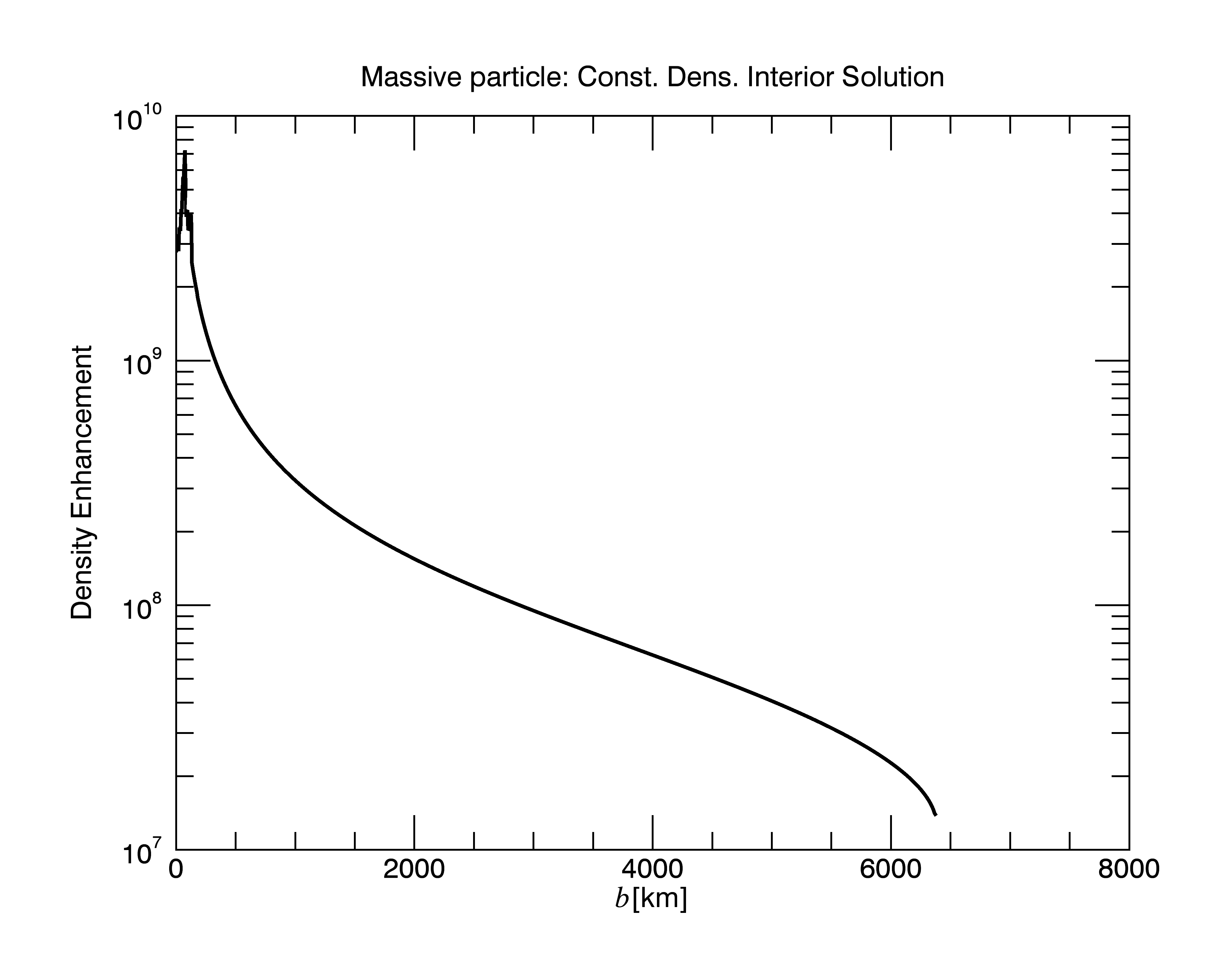}}
\caption{Density enhancements of   massive particles streaming through a constant density Earth as a function of impact parameter $b$, assuming $\rd=1$.  Note the large spike at small impact parameters where the focal points are effectively independent of the impact parameter with corrections of order $b^2/\re^2$.  The density enhancement peak for a constant density planet can be located: for this Earth model, it occurs at impact parameter $b=55$~km. }\label{massiveConstMag}
\end{figure}
\begin{figure}
\resizebox{9cm}{!}{\includegraphics*{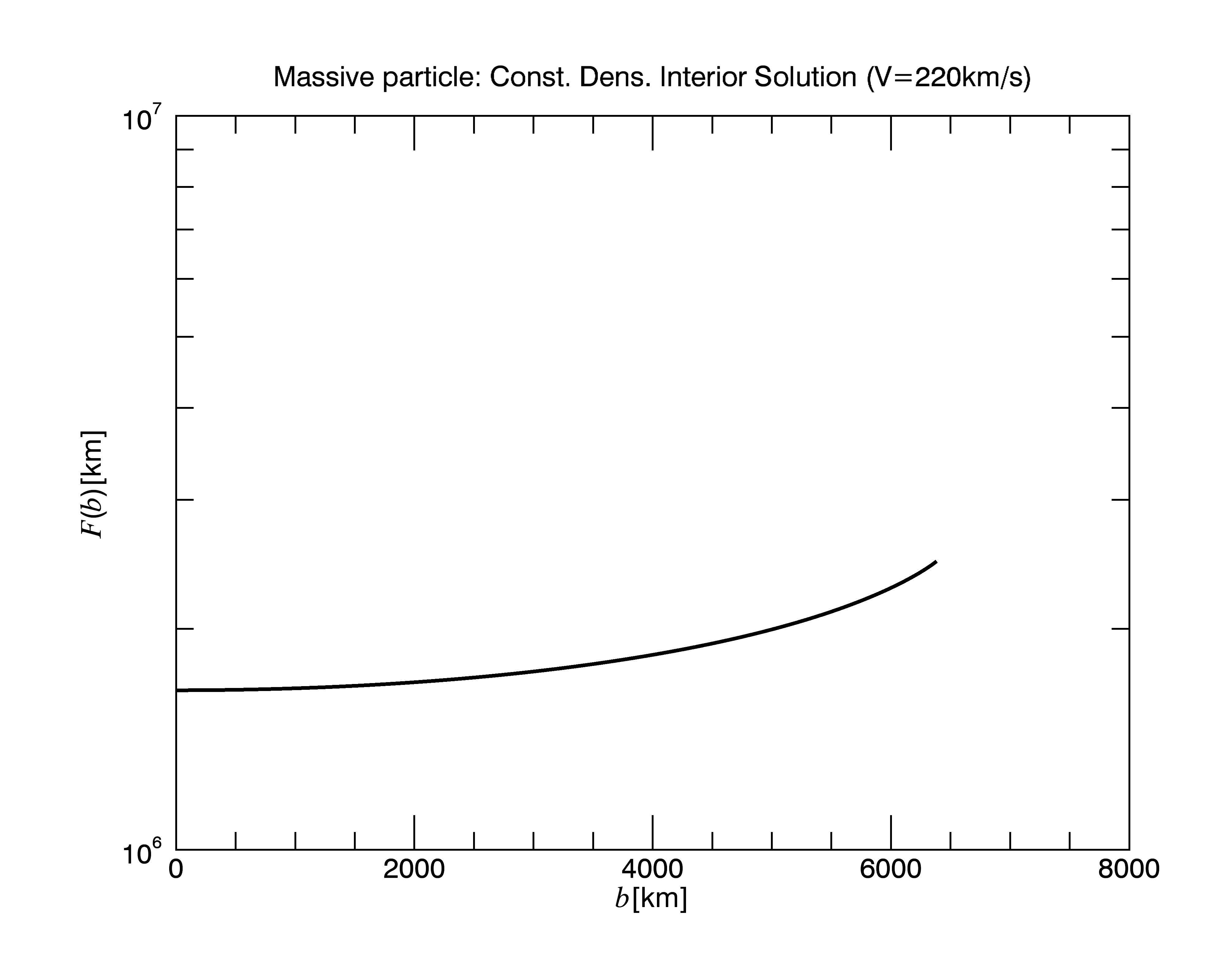}}
\caption{Focal points of   massive particles streaming through a constant density Earth as a function of impact parameter $b$.  The root of a constant density hair stemming from particles moving at 220~km/s is at a distance similar to L2 (with a different direction) and is relatively accessible.}\label{massiveConstFP}
\end{figure}
\begin{figure}
\resizebox{9cm}{!}{\includegraphics*{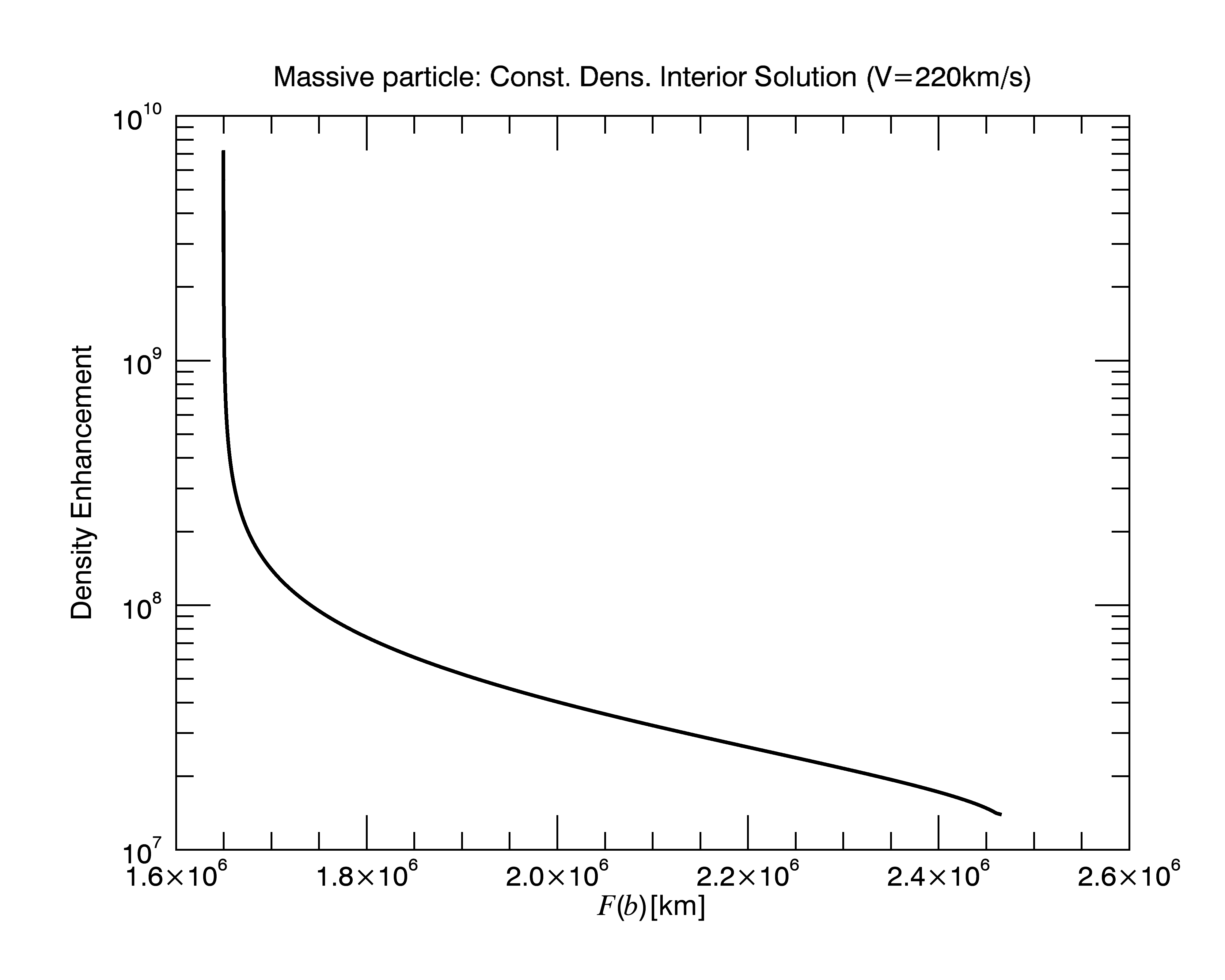}}
\caption{Density enhancements of   massive particles streaming through a constant density Earth as a function of focal point distance.  The enhancement curve is effectively vertical at the root of the hair suggesting a dense, point-like structure trailed by a sparser tail.  Note that this "sparse tail" is still highly dense as compared to the ambient density of CDM.}\label{massiveConstMagFP}
\end{figure}

\section{Fast Accurate Integrand Renormalization (FAIR) Algorithm}
In order to calculate the geodesics for realistic planetary radial density models, it is necessary to quickly and accurately perform integrals  of the form
\beq
\int_{ \phi_\text{i} }^{ \phi_\text{f} }  \text{d}\phi =\int_{R_{\text{body}}}^{ r_\text{f} }  \frac{     vb/(cr^2)\text{d}r   }{\sqrt{   \left[1-\frac{\rs}{r}\frac{M(r)}{M_{\text{E}}}\right]  \left[ \left(    1 + \frac{v^2}{c^2}  \right) e^{-2\Phi/c^2} \!- 1\! - \frac{v^2b^2}{c^2 r^2}   \right] } }
\eeq
for a general potential $\Phi$ and where $\phi_\text{i}$ is the initial angle determined by the exterior solution of the geodesic equation and where $r_\text{f}$ is the radius of closest approach at $\phi_\text{f}$.  Since the integrand is singular at $r=r_\text{f}$, $\phi_\text{f}$ is difficult to evaluate numerically and the deflection angle may be hard to pin down to sufficient precision.  

Early  in this work, Runge-Kutta methods were  used to solve the geodesic equations, but they were found to converge  too slowly.  Other numerical methods found in the literature~\citep{press2007numerical} were tried with unsatisfactory results because $\phi_\text{f}$ could not be calculated accurately enough near the caustic of the focal point.  Indeed, the density enhancement depends on the derivative of the impact parameter with respect to the focal point (see \Eq{mageq}) which is huge for small impact parameters.  To get around this problem, a method was developed to perform the integral in a quasi-analytic fashion that relied on the physical intuition that the precise details of the density should have little effect on the overall shape of the geodesics.  In other words, the precise form of the density on a scale of a few meters should  not significantly impact the interior solution of  a large body with a radius of thousands of kilometers.  Additionally, none of the planetary models claim to be accurate at small scales, which leaves us with the flexibility to choose a convenient density function that can be solved analytically on the small radial scale\footnote{The word renormalization refers to this flexibility since renormalization in  quantum field theory is effectively a modification of short scale physics.  For example, propagator functions can be modified by multiplicative functions that make integrands analytic at their short scale singularity.   }.  This algorithm also allows a test of the robustness of the qualitative features of the density enhancement, like the sharp peak at low impact parameters.

\begin{figure}
\resizebox{9cm}{!}{\includegraphics*{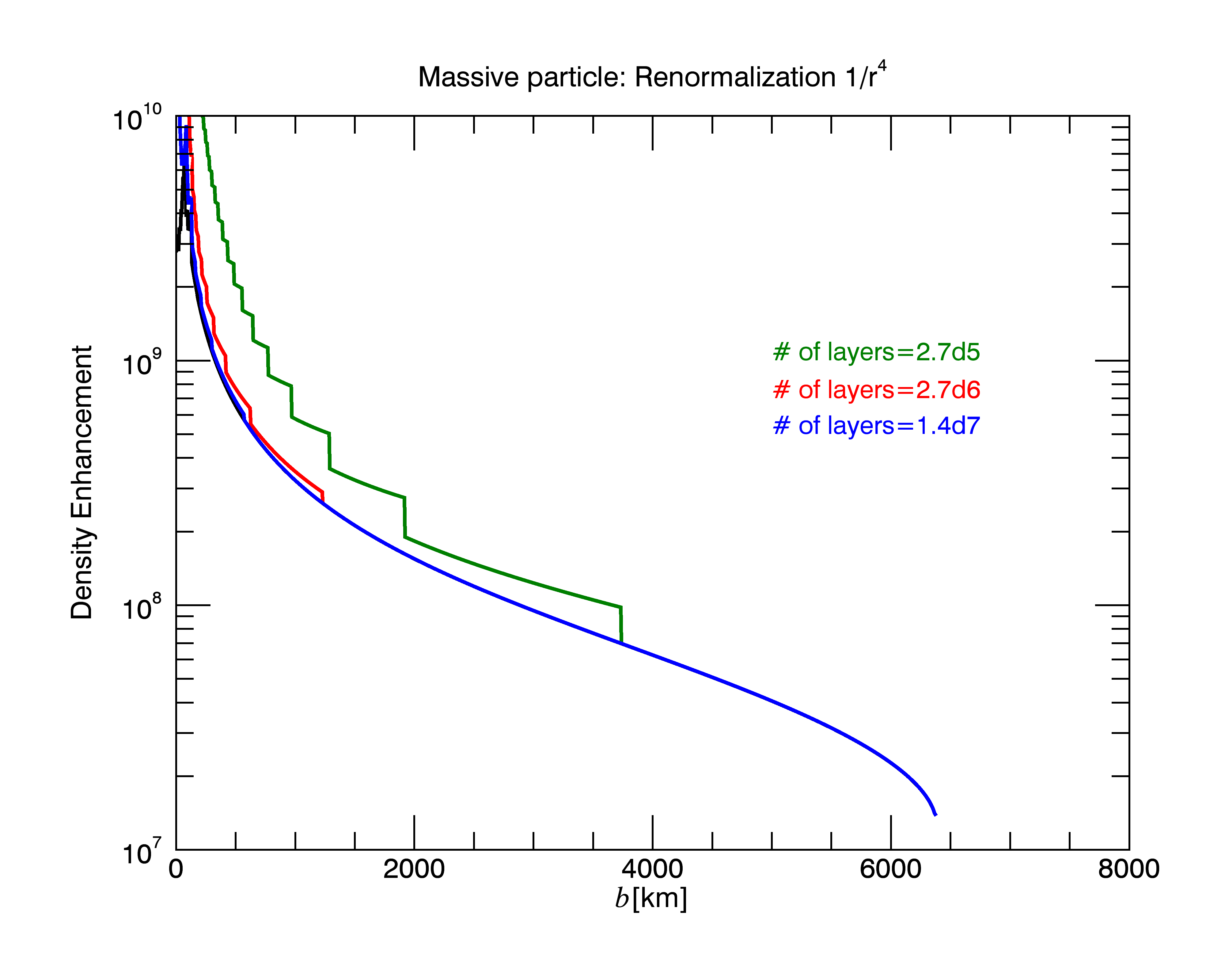}}
\caption{Density enhancement results from using a $r^{-4}$ radial dependence within the density segments.  The black analytic solution   is covered by the blue line of the FAIR algorithm with $10^7$ layers. }\label{magr4}
\end{figure}

\begin{figure}
\resizebox{9cm}{!}{\includegraphics*{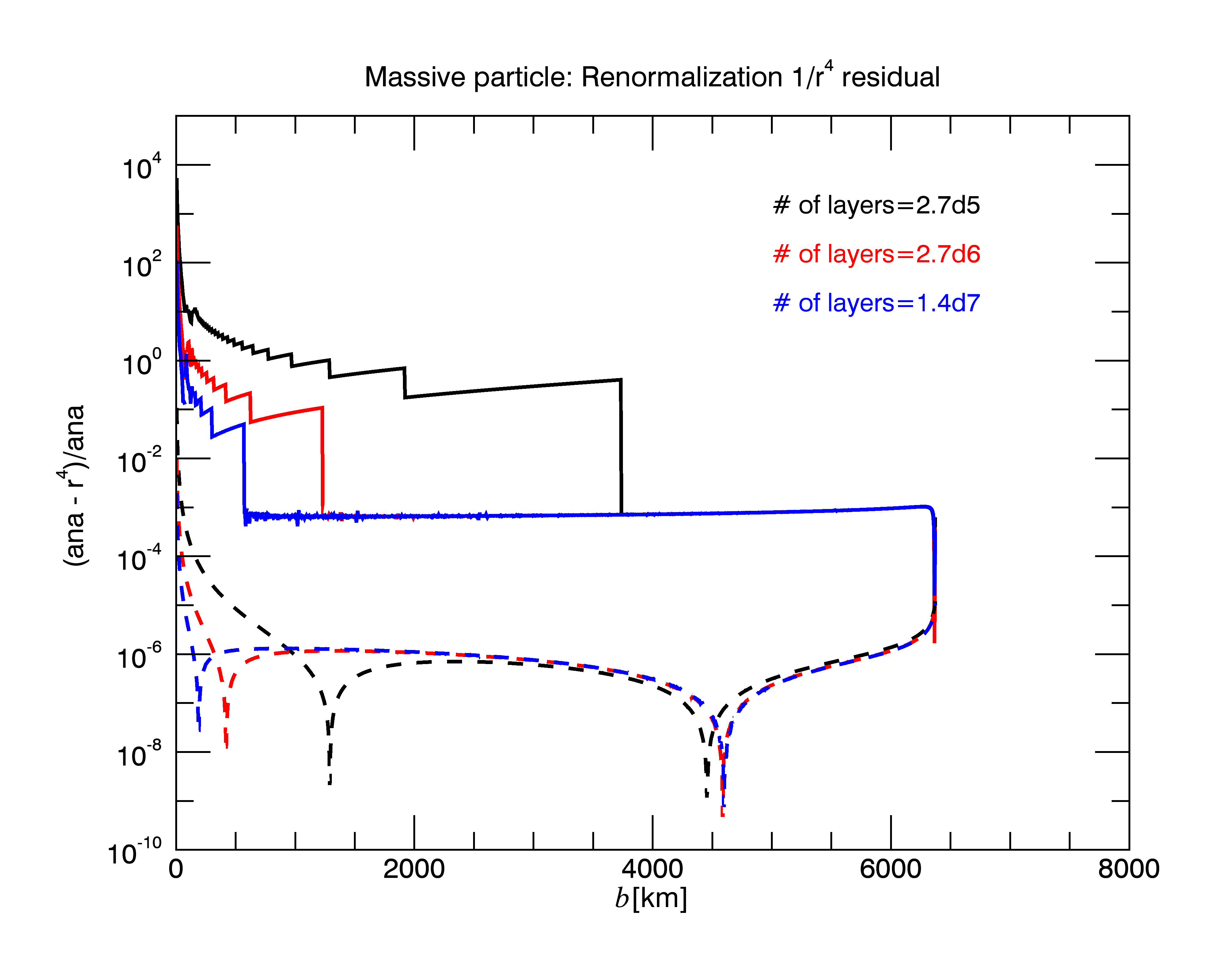}}
\caption{Density enhancement (solid) and focal point (dash) residuals between the constant density analytic and interior solutions solved using a $r^{-4}$ density renormalization of a constant density Earth for increasing number of layers. Using $10^7$ steps, the $r^{-4}$ solution is seen to be accurate to about $b=500$~km which is orders of magnitudes better than the RK solution (not shown) for the same number of steps.}\label{residualr4}
\end{figure}

Assuming the prior existence of a radial density model for a particular body, the basic FAIR algorithm works as follows:
\begin{itemize}
\item Split the integrand into $N$ segments of length much smaller than the radius of the body.
\item Create a density profile within each segment that is analytically solvable.  Equate the average density of the density profile in that segment to the known, average, model value within that segment.
\item Starting with the known $\phi_{i+1}=\phi_\text{i}$ from the exterior geodesics equation, use the analytic solution of the integral inside that segment to calculate the next value of the angle $\phi_i$.
\item Do this until you reach the segment that contains the value of $r=r_\text{f}$ at some layer $j$.  At that point, calculate $\phi_\text{f}$ with
\beq
\phi_\text{f} = \phi_{j+1} - (\text{analytic solution evaluated at }r=r_{j+1})
\eeq
\end{itemize}
The FAIR algorithm was verified numerically by comparing the  analytic solution of the constant density Earth with a highly unphysical case where each layer had  a short scale  radial dependence of $1/r^4$ 
\beq\label{dp4}
\rho = \sum_{i=0}^{N_{\text{layers}}-1} \frac{\rho_i}{r^4}
\eeq
subject to the average-density-equality condition
\beq
\rho_i = \frac{\bar{\rho}_i}{3} \frac{  r_{i+1}^3 - r_i^3  }{ 1/r_i - 1/r_{i+1}  }~,
\eeq
where $\bar{\rho}_i$ is the density value extracted from a standard  model at the $i^{\text{th}}$ layer.  In the constant density case, $\bar{\rho}_i=\bar{\rho}$ for all $i$'s.  After a bit of algebra, the solution for each layer with the density profile of \Eq{dp4} is found to be
\beq
\phi_i &=& \phi_{i+1} - \text{acos}\left[  \frac{  \frac{2\gamma_i^2}{\beta_i}   u_{i+1}  -1    }{  \sqrt{   \frac{4\gamma_i^2 \alpha_i   }{ \beta_i^2   }   + 1 }       }   \right]   +   \text{acos}\left[  \frac{  \frac{2\gamma_i^2}{\beta_i}   u_{i}  -1    }{  \sqrt{   \frac{4\gamma_i^2 \alpha_i   }{ \beta_i^2   }   + 1 }       }   \right] \\
\alpha_i &\equiv& \alpha_{i+1}  - \frac{4\pi G}{b^2 v^2} u_{i+1} ( \rho_i - \rho_{i+1} ) \\
\beta_i &\equiv& \frac{8\pi G}{b^2 v^2} \left(  \frac{M_{\text{body}}}{4\pi} + \rho_i u_{i+1} \right)  \\
\gamma_i^2 &\equiv& \frac{4\pi G}{b^2 v^2} \rho_i + 1 ~ ,
\eeq
where $u_i\equiv 1/r_i$ and $\phi_{i+1}$ was calculated in the previous step.  Comparing the results for the density enhancement of the analytic solution to the unphysical $r^{-4}$ in \Fig{magr4}, the agreement improved with the number of layers and was quite good  for $10^7$ layers demonstrating a  resolving robustness of the qualitative rapid magnification rise for decreasing impact parameter.  The residuals for both the density enhancement and focal point locations are provided in \Fig{residualr4} further showing how the FAIR algorithm tends to the right answer with increasing number of layers even for a highly unphysical case.  The $10^7$ layer case took about 10 minutes to run and demonstrates that even if your initial guess for the intra-layer density radial dependence is very off, you can still obtain a very good result compared to the true solution rather quickly.

The simulated results appearing below for PREM and J11-4a were obtained with the {\it geodesolver} code based on the above algorithm but with constant intra-density  layers   more in line with physical expectations.  The iteration equations of {\it geodesolver} are
\beq\label{geodeq}
\left(  \frac{\text{d}u}{\text{d}\phi} \right)^2 &=& \frac{1}{u^2} \left[ a_0 + a_1 u^2 + a_2 u^3 - u^4 \right] \\
a_0 &\equiv& -\frac{4\pi G}{3v^2b^2} \bar{\rho}_i \\
a_1 &\equiv& \frac{1}{b^2} + \frac{1}{v^2b^2}\left( |V_{i+1}| + 4\pi G\bar{\rho}_i r^2_{i+1} \right) \\
a_2 &\equiv& \frac{2G}{v^2b^2} \left( \frac{M_{i-1}}{r_i}  - \frac{4\pi}{3}\bar{\rho}_ir_i^2   \right) \\
M_{i-1} &\equiv& M_i -  \frac{4\pi}{3}\bar{\rho}_i\left(r_i^3-r_{i-1}^3\right) \\
V_{i+1} &\equiv& V_{i+2} + 4\pi G\bar{\rho}_{i+1} \left(r_{i+2}^2-r_{i+1}^2\right)
\eeq
where the starting values of $r$, $M$, and $V$ are the radius of the planet, the mass of the planet, and zero, respectively; the $\bar{\rho}$'s are given by the physical planetary model.  To solve \Eq{geodeq}, it is best to remove the $r^3$ term with the following trick:
\beq
u^3 &\cong& u_0^3 + \delta u^3 \\
&\cong& u_0^3 + 3u^2 (u-u_0)
\eeq 
and solve for $u^3$.  \Eq{geodeq} now has the same form as \Eq{constEq} and can be solved in the same manner.

\begin{figure}
\resizebox{9cm}{!}{\includegraphics*{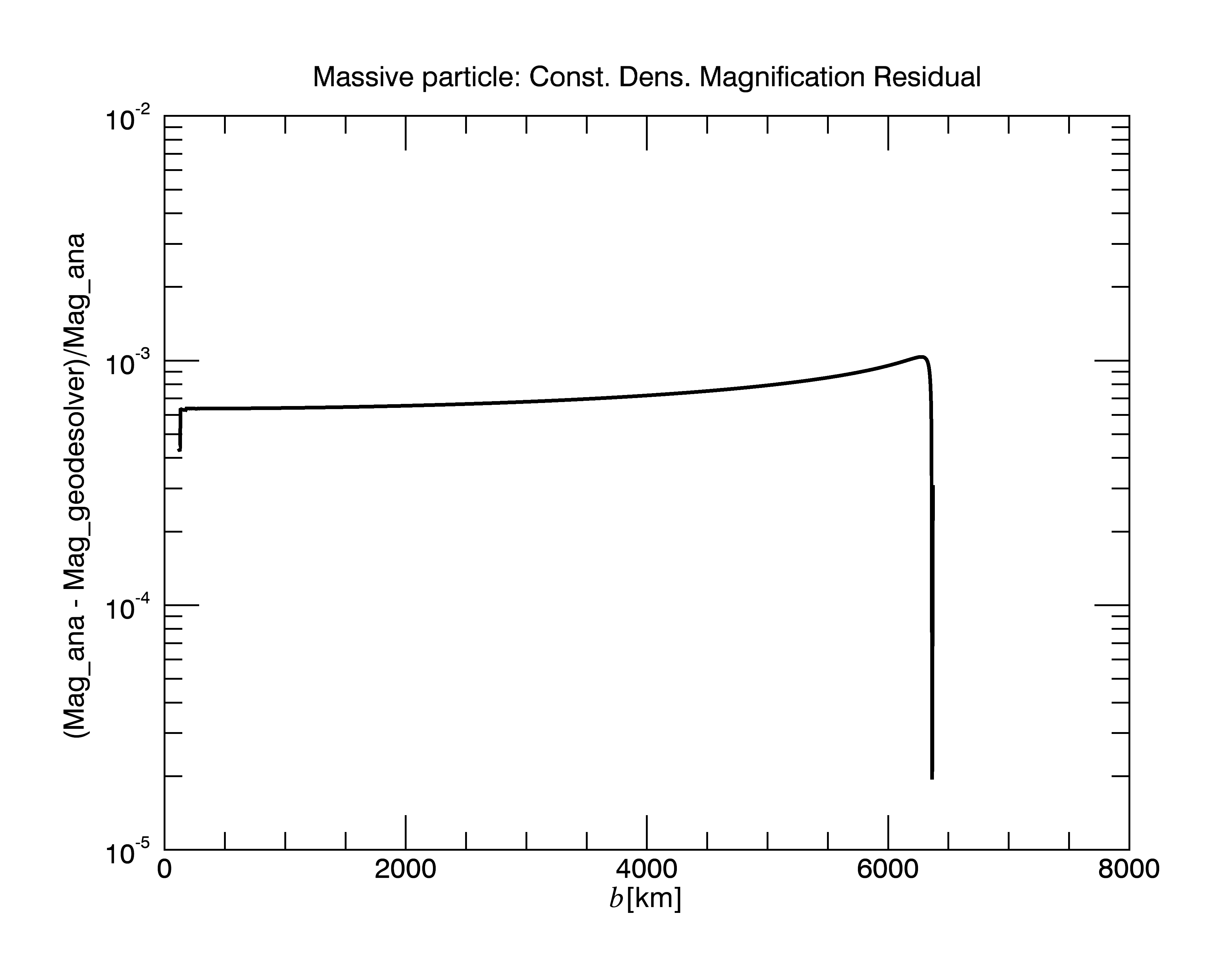}}
\caption{Density enhancement residual between the analytic and {\it geodeslover} interior solutions of a constant density Earth for $3\times10^5$ layers. }\label{geodemag}
\end{figure}

\begin{figure}
\resizebox{9cm}{!}{\includegraphics*{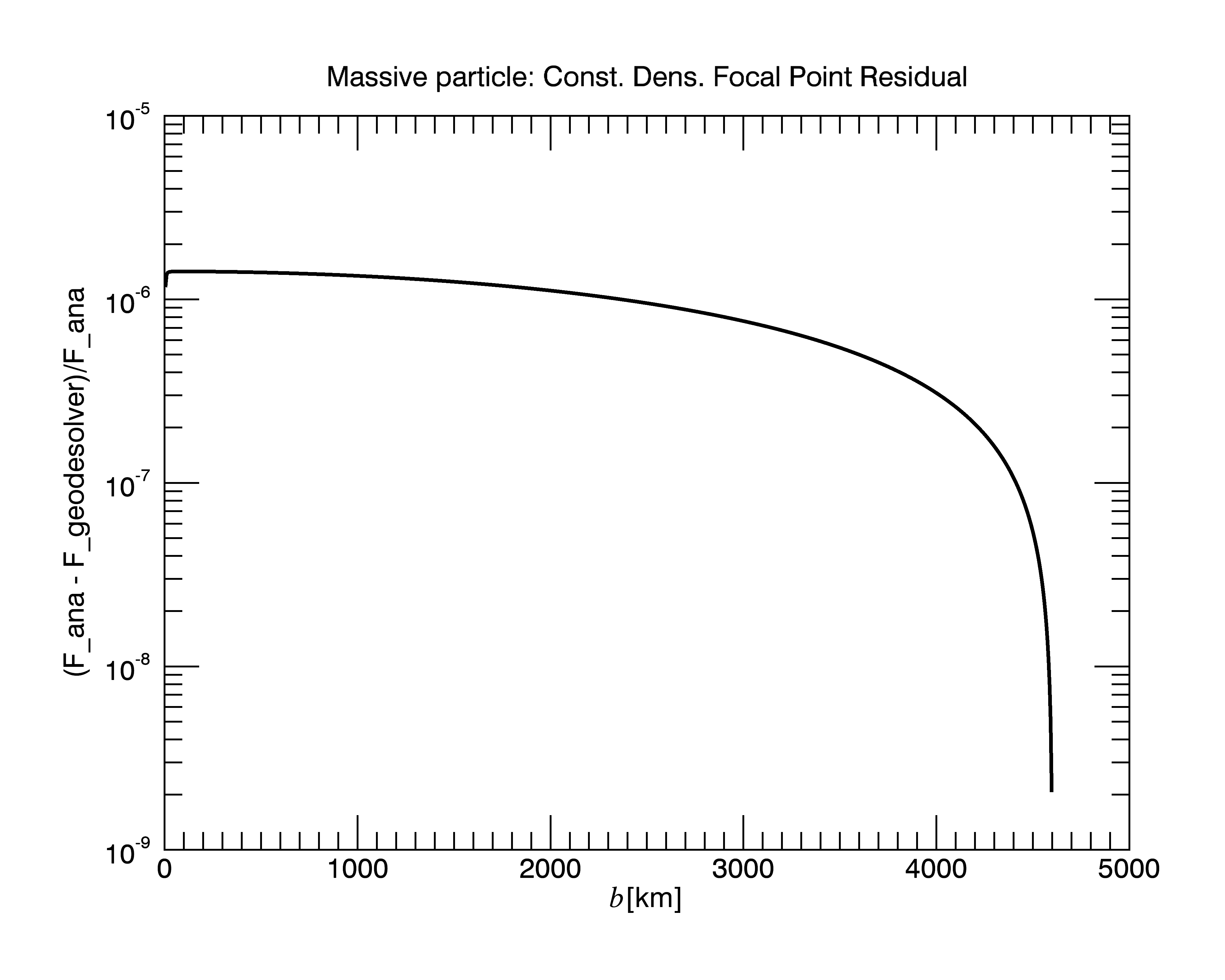}}
\caption{Focal point residual between the analytic and {\it geodeslover} interior solutions of a constant density Earth. }\label{geodefp}
\end{figure}

The agreement between {\it geodesolver} and the analytic solution for the constant density Earth is excellent: using only $3\times10^5$ layers, the density enhancements agree with the constant density analytical case at the 0.1\% (\Fig{geodemag}) level while the focal points match  to 6 significant digits (\Fig{geodefp}),  including  near the critical  sharp density enhancement peak at small impact parameter.  At $10^5$ layers, running {\it geodesolver} takes only a few seconds to calculate each point.

\section{Interior Solutions: Preliminary Reference Earth Model and J11-4a Jupiter}
\begin{figure}
\resizebox{9cm}{!}{\includegraphics*{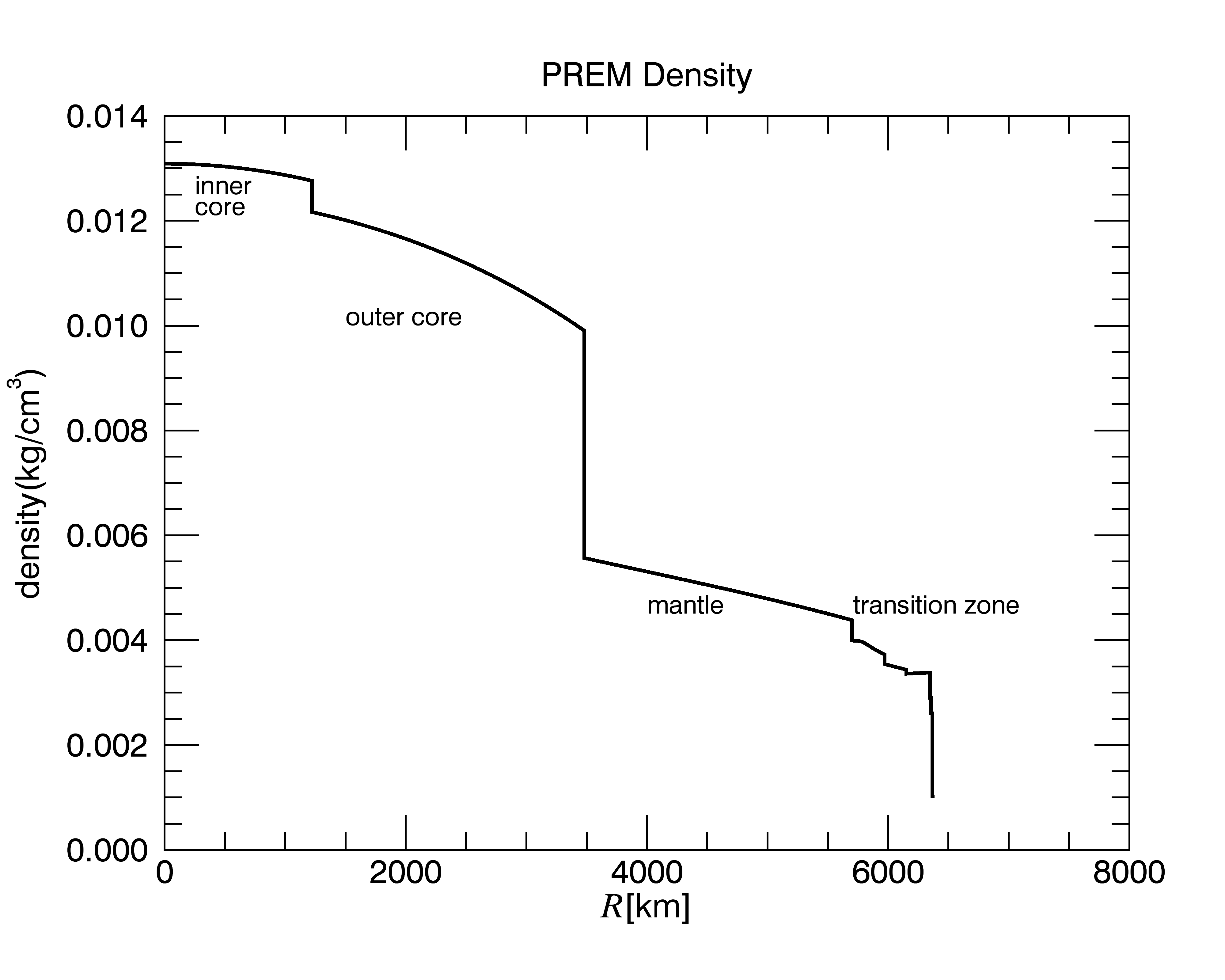}}
\caption{Preliminary Reference Earth Model radial density.}\label{premdp}
\end{figure}

\begin{figure}
\resizebox{9cm}{!}{\includegraphics*{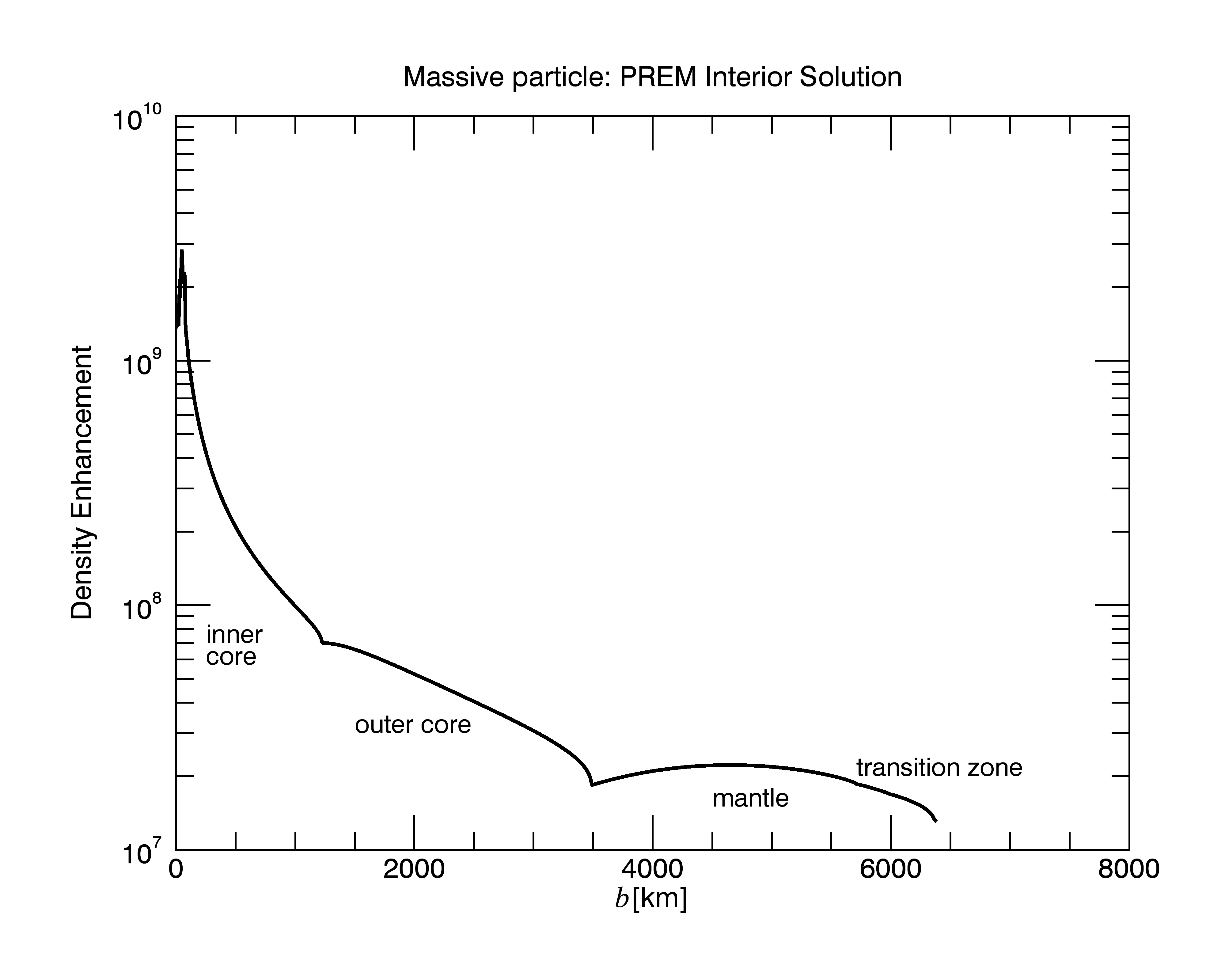}}
\caption{Density enhancement of  massive particles streaming through a PREM model Earth as a function of their impact parameter.   Earth's layers are clearly visible showing that a hair could be a  highly accurate probe of Earth's interior.}\label{premmag}
\end{figure}

\begin{figure}
\resizebox{9cm}{!}{\includegraphics*{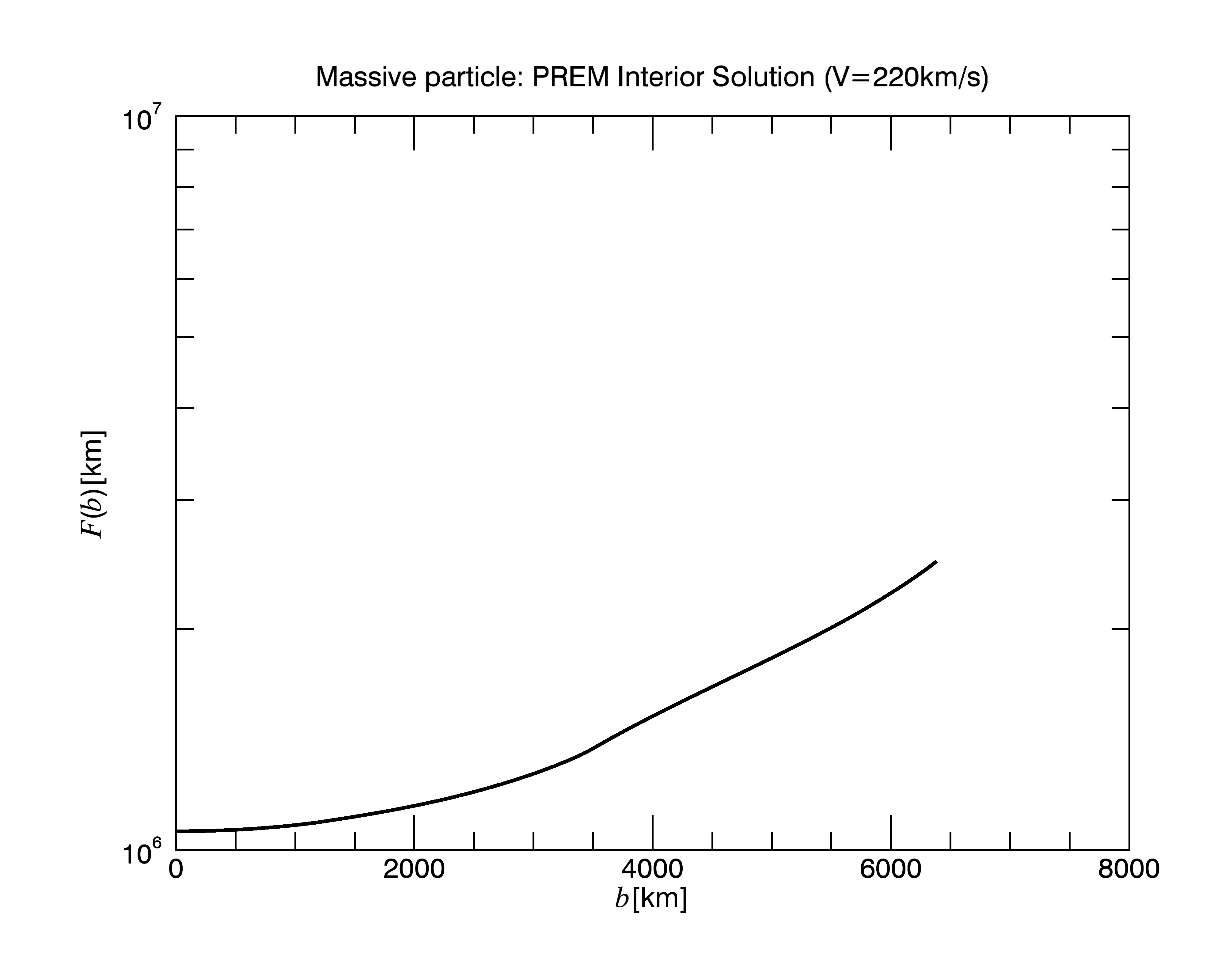}}
\caption{Focal points of  massive particles streaming through a PREM model Earth as a function of impact parameter $b$.  The root of the hair is located at a distance of 165~$\re$.}\label{fpEarth}
\end{figure}

\begin{figure}
\resizebox{9cm}{!}{\includegraphics*{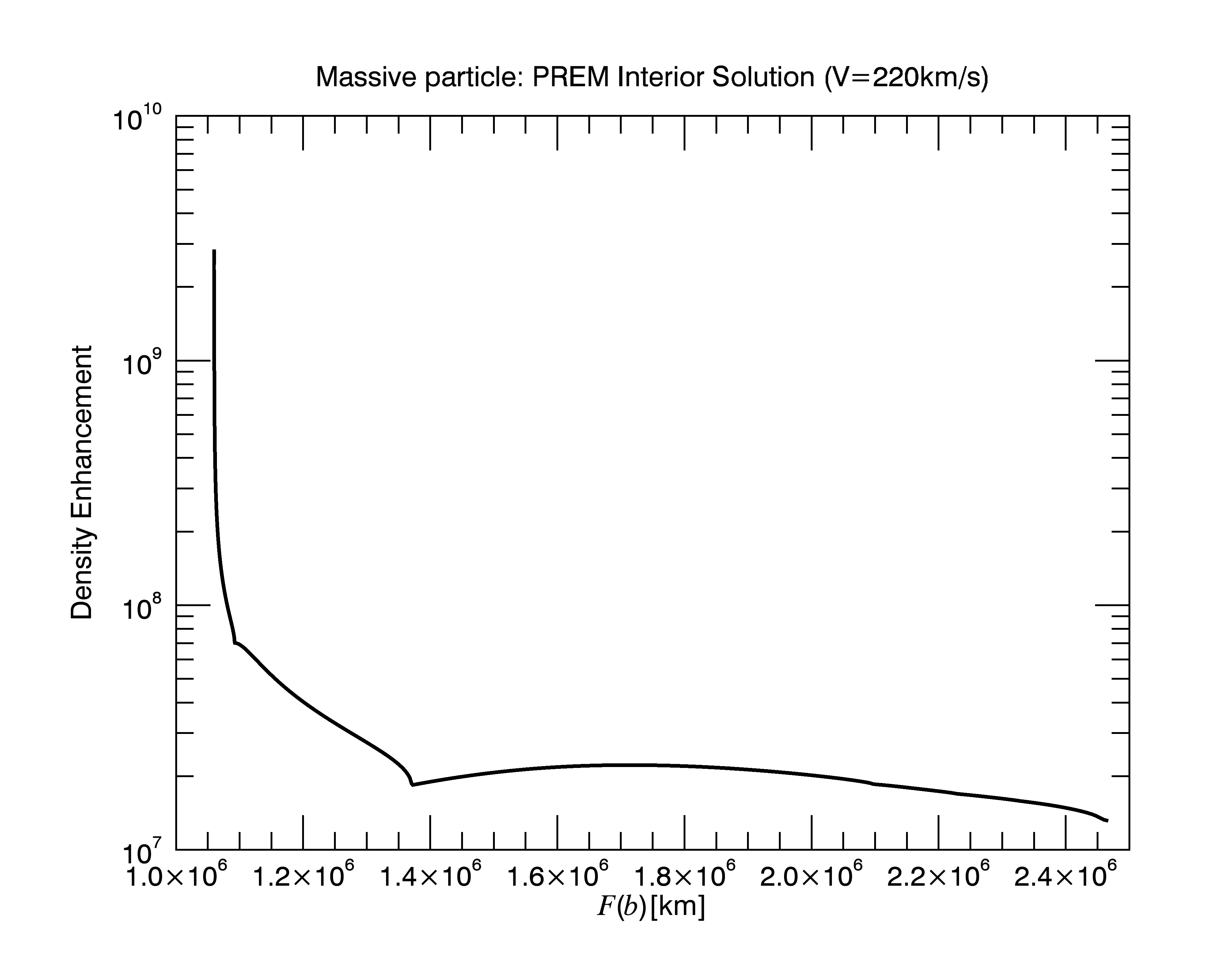}}
\caption{Density enhancements of  massive particles streaming through a PREM model Earth as a function of focal point distance.}\label{magfpEarth}
\end{figure}

{\it geodesolver} was used on  realistic planetary models for Earth (PREM~\citep{1981PEPI...25..297D}) and Jupiter (J11-4a~\citep{Nettelmann:2011kx}); these density profiles are plotted in Figs.~\ref{premdp} and \ref{j11dp} respectively.  The density enhancements and focal points for PREM  are plotted in Figs~\ref{premmag}, \ref{fpEarth}, and \ref{magfpEarth}  while Figs~\ref{j11mag}, \ref{j11fp}, and \ref{j11magfp} are the corresponding plots for J11-4a.  The density enhancement plots are particularly striking as they  clearly resolve the interior structure of the planets.  Additionally, they also exhibit huge density enhancements for small $b$'s with point-like root structures as in the constant density Earth model.  Other characteristics of the plots are tabulated in  Table~\ref{tabval}. The second column of Table~\ref{tabval} refers to the cutoff velocity of the hair, namely the velocity at which the root of the hair is below the surface of the planet.  This velocity constrains searches for hairs surrounding massive bodies  as  entire hairs from root to tip may find themselves inside a body for likely values of the velocity (e.g., the Sun).  That cutoff velocity can numerically be calculated using {\it geodesolver} but it is also easily obtained from the roots already calculated at $v=220$~km/s using the constant density result that the root distance grows quadratically with the velocity (\Eq{constfpeq}):
\beq
v_\text{c}=v\sqrt{\frac{R_\text{body}}{F_\text{r}}}
\eeq
The fact that relations derived for a constant density Earth are applicable to realistic planetary models is explained below.
\begin{figure}
\resizebox{9cm}{!}{\includegraphics*{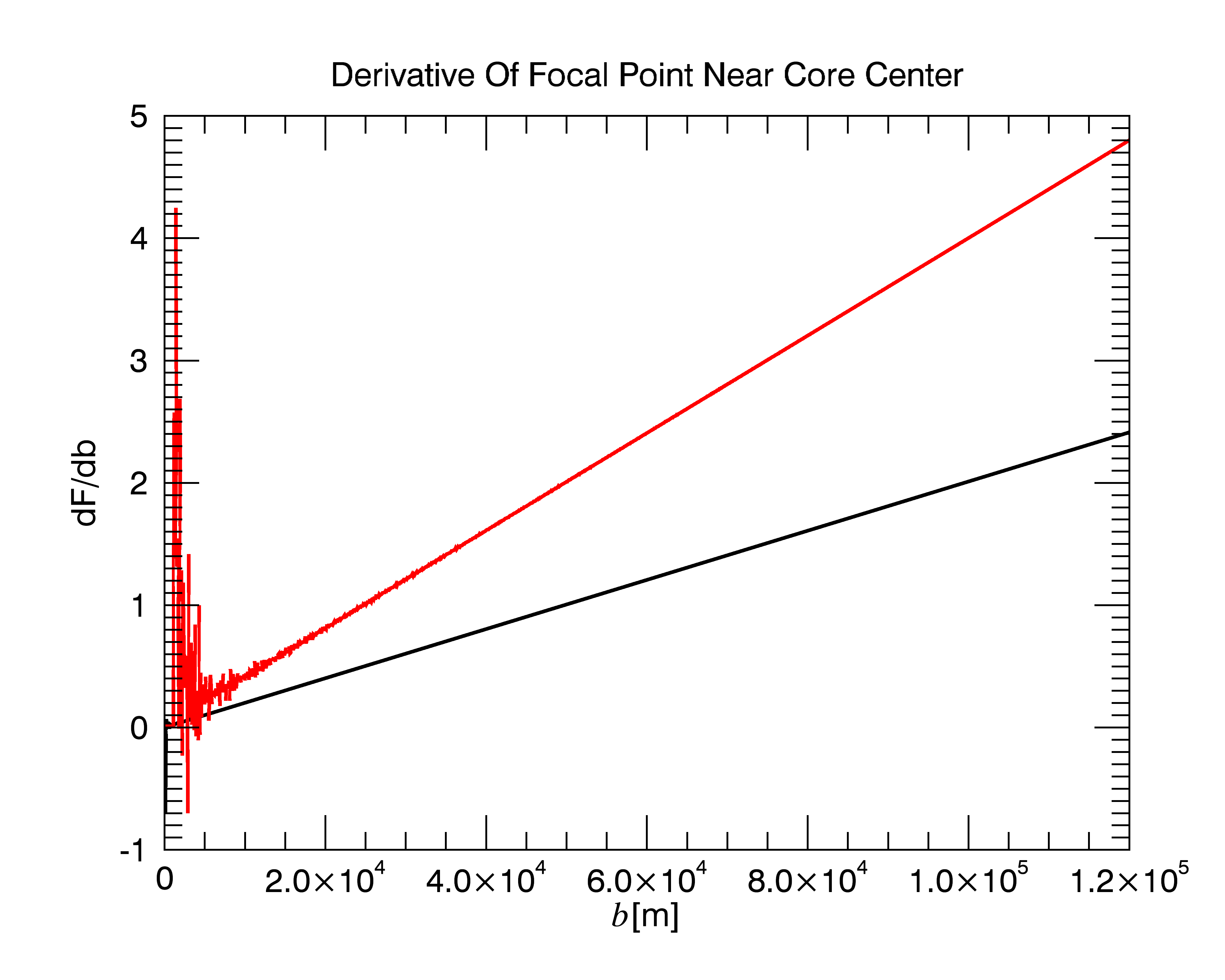}}
\caption{Derivative of the focal point with respect to the impact parameter $b$ for constant density (black curve) and PREM Earth (in red).  Near $b=0$, the derivative is dominated by numerical noise.}\label{derivplot}
\end{figure}
\begin{figure}
\resizebox{9cm}{!}{\includegraphics*{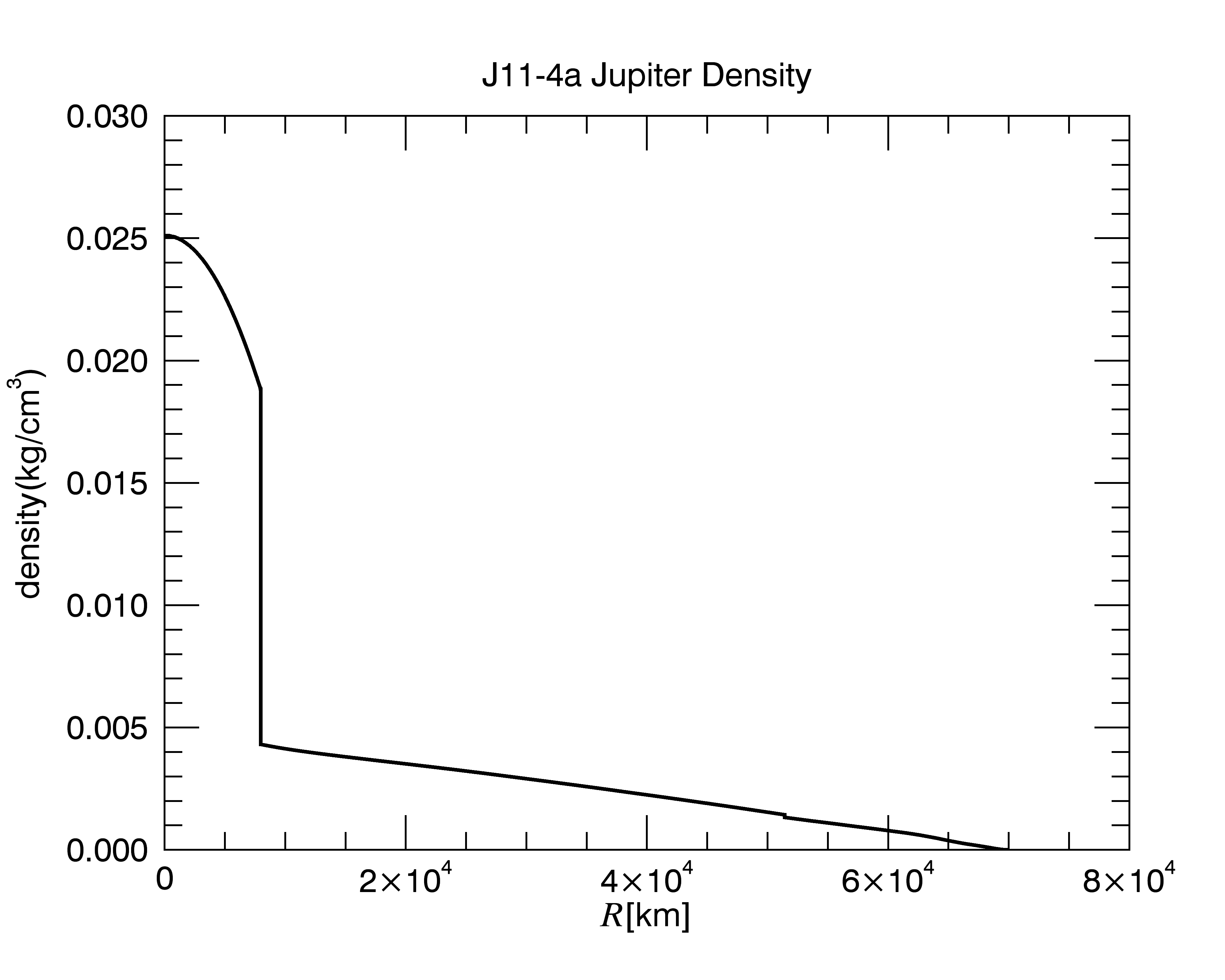}}
\caption{J11-4a Jupiter radial density.}\label{j11dp}
\end{figure}
\begin{figure}
\resizebox{9cm}{!}{\includegraphics*{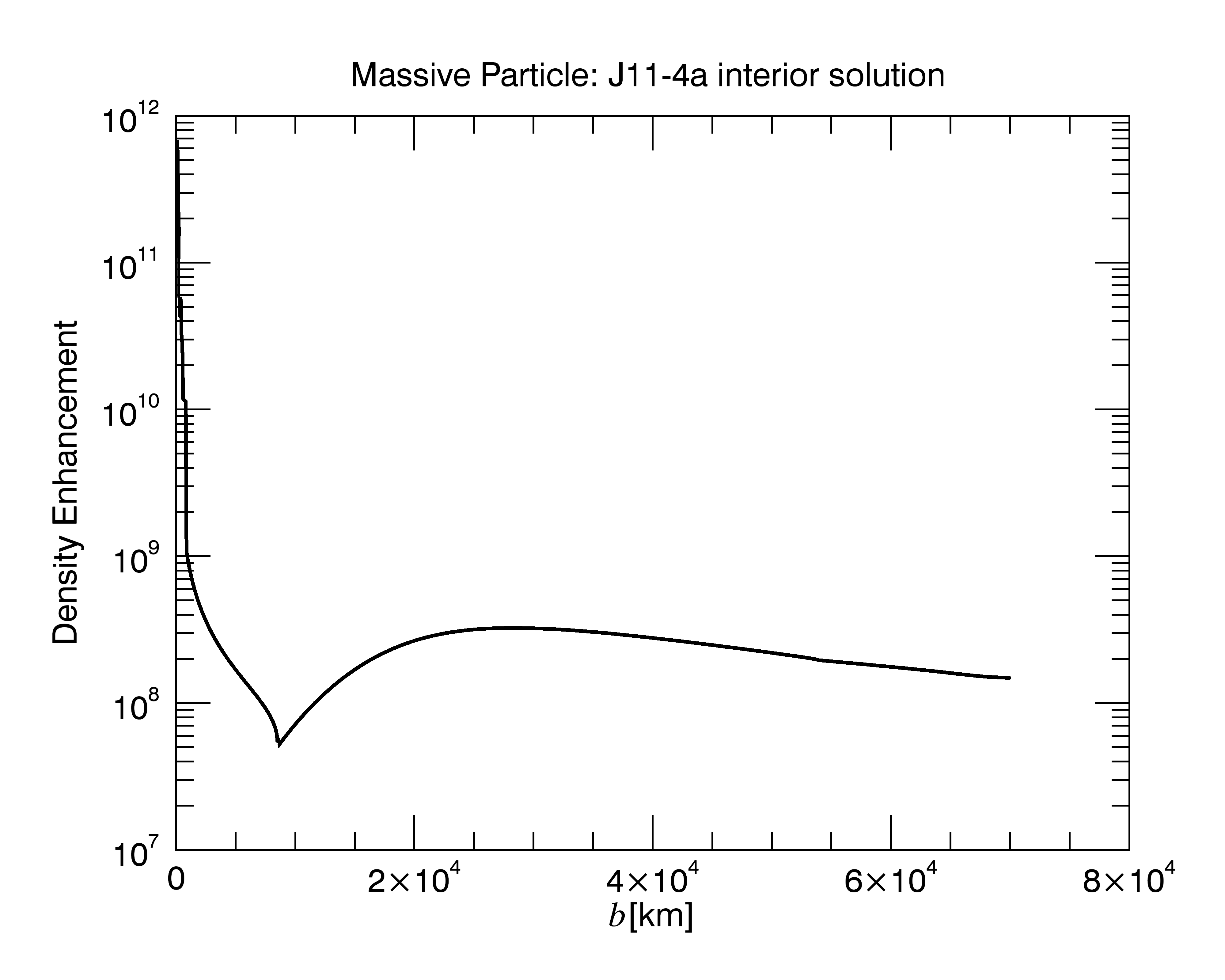}}
\caption{Density enhancement of  massive particles as a function of their impact parameter through Jupiter.  Once again, the interior structure is clearly visible in the hair density enhancement profile.}\label{j11mag}
\end{figure}
\begin{figure}
\resizebox{9cm}{!}{\includegraphics*{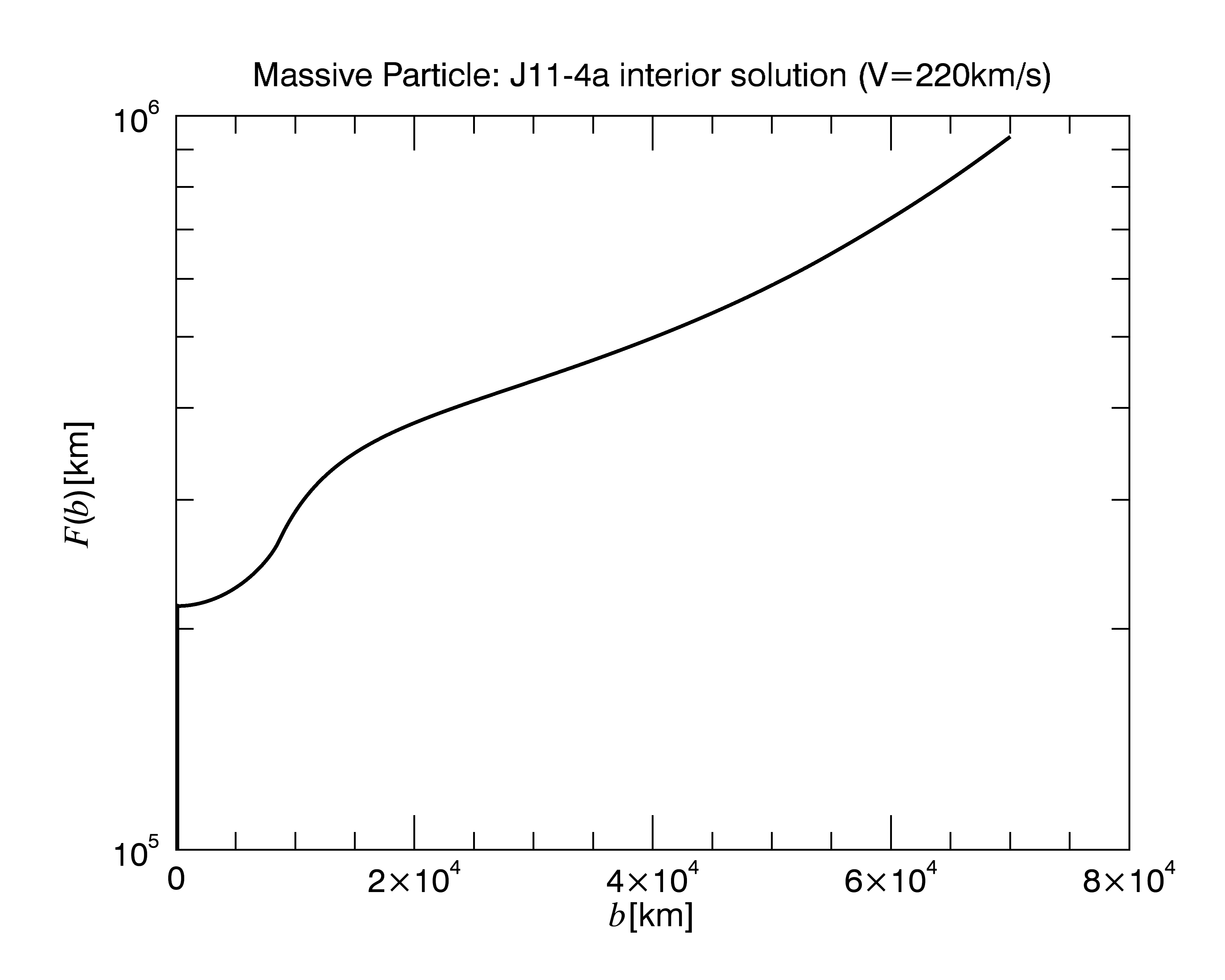}}
\caption{Focal points of   massive particles streaming at 220~km/s through a J11-4a Jupiter model as a function of impact parameter $b$.  The root of the hair is located at 3 Jupiter radius and any search seeking to take advantage of the large density enhancements of dark matter fluxes near Jupiter would also have to deal with the intense fields surrounding the gas giant.}\label{j11fp}
\end{figure}
\begin{figure}
\resizebox{9cm}{!}{\includegraphics*{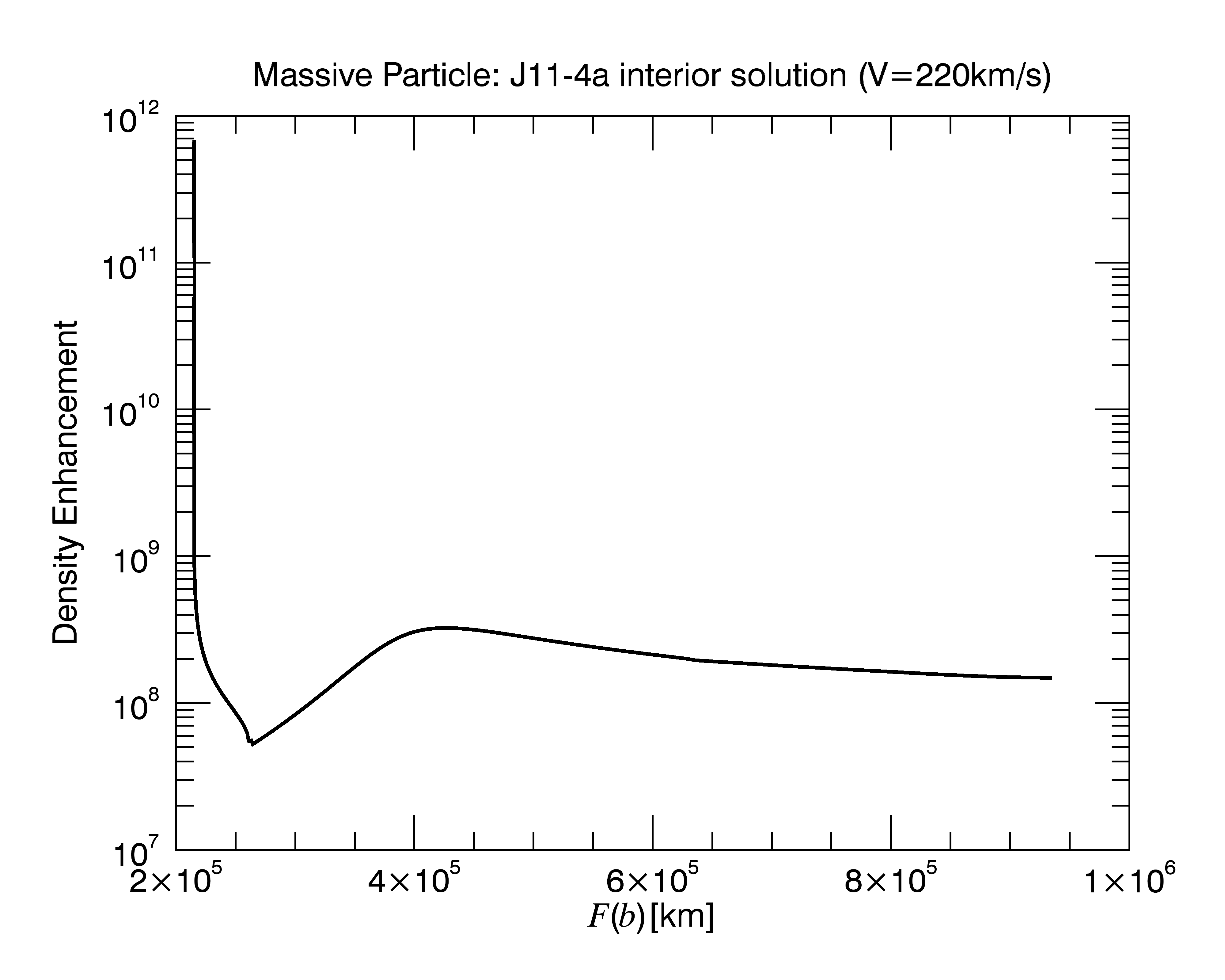}}
\caption{Density enhancements of   massive particles streaming at 220~km/s through a J11-4a Jupiter model as a function of focal point distance.}\label{j11magfp}
\end{figure}
To better understand the behavior of the focal points as $b\rightarrow0$, let's blow up the small impact parameter segment of the focal point plot in Fig.~\ref{fpEarth} and plot the derivative $[F(b_{i+1})-F(b_i)]/(b_{i+1}-b_i)$ as a function of $b$ in \Fig{derivplot}.  Thus, even for a realistic model such as the PREM, the derivative of the focal point is linear with the impact parameter as $b\rightarrow0$.   This implies that the focal points are constants save for tiny corrections  of the order $b^2/R^2_{\text{body}}$.  The fact that this result generally holds for arbitrary radial density profiles   can be understood by rewriting \Eq{massgeoeq}
\beq \label{b2exp}
\text{d}\phi=\frac{     vb\text{d}r   }{ cr^2  \left[1-\frac{\rs}{r}\frac{M(r)}{M_{\text{E}}}\right]^{0.5}   \left[2\left|\frac{\Phi}{c^2}\right| + \frac{v^2}{c^2}\left(1-\frac{b^2}{r^2}\right)   \right]^{0.5} }
\eeq
and expanding in powers of $b$.  Note that the singularity of the integrand is not within the integration interval, but at the lower limit where the integrated result is zero since this is the  turning point where the radial distance is minimized, and about which the particle trajectory is symmetric.  Thus, similar huge density enhancements of coherent particles exiting the core can be expected  for any compact body in the Schwarzschild metric, as was  verified analytically for the constant density Earth and numerically for the PREM and J11-4a.  Quite generally, the focal points corresponding to impact parameters near the core of  a Schwarzschild compact body can be parameterized by
\beq
F(b)=F_{\text{r}}+\frac{F_{\text{r}}^\prime}{2} b^2
\eeq
In Table~\ref{tabval}, these parameters  are given in columns 3 and 4 for PREM and J11-4a.
\begin{table}
\begin{tabular}{|c|c|c|c|c|}
  \hline
  Body & $v_{\text{c}}$~(km/s) & $F_{\text{r}}(\text{m})$ & $F_{\text{r}}^\prime (\text{m}^{-1})$  & PM \\
  \hline\hline
  ECD & 14  & $\xi\re^2/3$ &  $\xi / 6$   &  $\left[ \frac{ (4\rd\re^2)^{\frac{1}{3} } }{\rd} \right]^2$ \\
  \hline
  PREM & 18 & $10^{9}$ &  $4\times10^{-5}$  & $3.1\times10^{9}$\\
  \hline
  J11-4a & 125 & $2.15\times 10^{8}$ & $1.3\times10^{-6}$  & $6.6\times10^{11}$ \\
  \hline
\end{tabular}
\caption{Values for different planetary models: ECD are the values for an Earth of constant density, PREM is the Preliminary Reference Earth Model and J11-4a is a realistic model for Jupiter.  $v_\text{c}$ is the cutoff velocity below which the root of the hair is below the surface of the planet.  $F_\text{r}$ is the focal point location of the root of the hair for $v=220$~km/s, while $F_{\text{r}}^\prime(\text{m}^{-1})$ parametrizes its $b^2$ dependence (see text).  PM is the peak density enhancement at the root for each case.  Finally, note that for PREM, the greatest density enhancement occurs for $b_\text{peak}$=48~km, similar to the result analytically derived for the ECD model (\Eq{bpeak}) and $b_\text{peak}$=1133~km for Jupiter.}\label{tabval}
\end{table}

\section{Discussion}

Key aspects of direct dark matter detection experiments  that are not empirically constrained is the incident dark matter flux on Earth and  the resulting lack of  knowledge of the velocity distribution of the CDM.  Traditionally, the gravitationally bound CDM has been modeled as an ideal gas with a Maxwellian velocity distribution
\beq
f_{\text{CDM}}(v) = \frac{4N}{\sqrt{\pi}\vs} \left(\frac{v}{\vs}\right)^2 \text{exp}\left( -\frac{v^2}{\vs^2} \right), ~~ v<\vesc,
\eeq
where $N$ is a normalization constant to ensure that the integral over the distribution results in the local CDM density, $\vs$ is the Sun's velocity with respect to a CDM halo taken to be stationary, and $\vesc$ is the escape velocity of the CDM from the galaxy.  This isotropic, analytically smooth velocity distribution is unrealistic since the CDM  had tiny primordial velocity dispersion at the last scattering surface.  It only acquired angular momentum and velocity dispersion through their interactions with gravitational wells and tidal forces.  The fraction of CDM that would have been thermalized and acquired an approximate Maxwellian velocity distributions are those that would have gone through many orbits about the Milky Way.  There will also be a fraction of CDM particles that has undergone fewer orbits such that the velocity distributions has peaks~\citep{Sikivie:1995dp}.

A number of realistic theoretical and numerical calculations have been performed to try and understand the velocity distribution in the halo ~\citep{Springel:2008cc,Hansen:2008ek,Vogelsberger:2008qb}.  In particular,  comprehensive and realistic fine-grained simulations~\citep{Vogelsberger:2010gd,Springel:2008cc,Vogelsberger:2008qb}  of the geodesics followed by the CDM produce a distribution of finely parsed streams with discrete velocities that can be treated effectively as smooth for experiments on Earth's surface.   There are  millions of streams flowing through our solar system  with $\rho_{\text{s}}>10^{-7}\langle\rho\rangle$, including at least a few streams with at least $10^{-3}\langle\rho\rangle$~\citep{Vogelsberger:2010gd}.  Detecting  hairs stemming from these streams would be  a  huge scientific windfall as they would offer us 
\begin{itemize}
\item a nearby, extraordinarily  intense source of cold dark matter particles accessible to space detectors studying the nature and interactions of CDM;
\item our only way to directly explore the local fine-grained stream structure, a prediction of $\Lambda\text{CDM}$;
\item a direct and detailed look at the radial density distribution of any planet or moon, since determining the location of a single hair near Earth or Jupiter would immediately provide the location of that  same hair extending from any other body in the solar system. The densest, fine-grained dark matter streams, that have undergone the fewest orbits, can be expected to be very smooth on length scales similar to the solar system. Indeed, the original, unclustered, dark matter flows that stretched and folded into fine-grained dark matter streams, would have been much larger than a solar system~\cite{Natarajan:2007tk}. The original dark matter stream would have had a CDM density near the cosmological average, but the subsequent stretching and folding explains why the particle density in a fine-grained stream is typically small: the total number of collisionless dark matter particles remained constant in an expanding volume.  The densest dark matter streams should therefore have the same velocity and density within the solar system resulting in hairs with near identical orientations at different planets\footnote{Any observed differences in the hairs between different planets would stem from  local corrections of the Schwarzschild metric due to nearby bodies.}.
\end{itemize}
Locating these hairs is likely to be challenging  but may be facilitated by the fact that the density enhancement is independent of the stream velocity, with minimum density enhancements of order $10^7$ for Earth and $10^8$ for Jupiter.  In addition, the detailed spherically symmetric density structure of the planetary layers does not impact the qualitative features of a hair such as a sharp rise at low impact parameters.

Although planets are not precisely described by a Schwarzschild metric, slight deformations should not  negate the basic arguments presented in this paper.  The reason is that planetary deformations  due to rotational motion do not break the mirror symmetry of the body across its center of gravity.  This mirror symmetry constrains the metric to depend on the square of the angular momentum, since a particle with angular momentum $\vec{L}$ and another with angular momentum $-\vec{L}$ will meet at the same focal point.  Hence, although planets are not precisely spherically symmetric,  the hairs will reflect those deformations in a proportional fashion such that a hair cross-section will simply represent a 2-dimensional image of the planet from that vantage point.  The issue of planetary deformation will be explored numerically in a future paper with a generalization of the FAIR algorithm.

Turning now to the parameters necessary for an estimate of the time and resources required for finding a hair, we have:
\begin{itemize}
\item $p$ represents the percentage of the sphere required for the search area.  This number needs to be calculated since the orientation of the hairs is biased by the galactic orbital motion of the solar system.  In addition, hairs with relatively large densities (say $\rho_{\text{s}}/\langle\rho\rangle>10^{-4}$) have likely gone through fewer orbits about the galaxy, and may have a different  orientation distribution from the low density streams.
\item $n_{\text{probe}}$ is the number of probes simultaneously conducting a search.  In estimating the time required to find a hair, $n_{\text{probe}}$ will be solved for as a indication of the feasibility of the search.
\item $\bar{T}_{\text{t}}$ is the average transit time of a probe through a hair.  This parameter is critical for the strength of the signal in the event the probe does fly through a hair.  If the transit time is too short for the sensitivity of the probe, the hair will not be detected.
\item $\bar{h}_{\text{w}}$, the average hair width crossed by a transiting probe.
\item $d_{\text{probe}}$ the distance from the body where the probe is concentrating its search.  In practice, one would expect  $d_{\text{probe}}=F_\text{r}$ to afford the best odds.
\item $N_{\text{s}}$ the number of streams for a particular density ratio $\rho_{\text{s}}/\langle\rho\rangle$.
\end{itemize}
The parameters above can be used to estimate the search time.  To solve for $n_{\text{probe}}$, that search time must be limited in some fashion.  If one requires the search time to be  less than a few years, the following equation follows
\beq
& &\text{search time}=\frac{p \bar{T}_{\text{t}} }{n_{\text{probe}}N_{\text{s}} }\frac{4\pi d_{\text{probe}}^2}{\bar{h}_{\text{w}}^2}< 10^8\text{s}\\
& &\zeta\equiv\frac{p \bar{T}_{\text{t}} }{n_{\text{probe}}N_{\text{s}} \bar{h}_{\text{w}}^2} < 10^{-11}~\frac{\text{s}}{\text{m}^2}\text{ for Earth or } \\
& &~~~~~~~~~~~~~~~~~~~~~~~~~<10^{-9}~\frac{\text{s}}{\text{m}^2} ~ \text{for Jupiter.}
\eeq
The parameter $\zeta$ depends on the details of the experiment, like the choice of target, the target size and the expected event rate.  In addition, although the transit time is constrained by the detector sensitivity, in practice it will also depend on the details of the probe orbit around the planetary body and how much fuel would be available to maneuver it around a desirable area of the focal sphere.  These experimental parameters will in turn determine the density enhancement required to detect a hair  during an average transit time, $\bar{T}_{\text{t}}$.  Taking as a requirement that a probe crossing a hair of width $\bar{h}_\text{w}$ see the same number of detection events as an Earth-bound mission would see in a year for an equivalent experimental setup, a constraint on $\bar{T}_{\text{t}}$ can be imposed
\beq\label{hw}
\frac{ \rho_{\text{s}}}{ \langle\rho\rangle }  \bar{M}\bar{T}_{\text{t}} \sim 10^7 \text{s} ~~\text{with}~~\frac{ \rho_{\text{s}}}{ \langle\rho\rangle }  \bar{M} >o
\eeq
where $\bar{M}$ is the average dark matter flux density enhancement experienced by the probe of radius $\rd$ as it transits across the width $\bar{h}_{\text{w}}$ of a hair and $o$ parametrizes the transit time and is therefore related to the probe perpendicular velocity through the hair cross-section.  Noting from Fig.~10 of Ref.~\cite{Vogelsberger:2010gd} that $(\rho_{\text{s}}/\langle\rho\rangle)  N_{\text{s}} \sim F(>\rho_{\text{s}})$, $\zeta$ can be rewritten
\beq
\zeta \sim \frac{p }{n_{\text{probe}}\bar{h}_{\text{w}}^2}  \frac{10^7\text{s}}{\bar{M}F(>\rho_{\text{s}})}
\eeq
where $F(>\rho_{\text{s}})$ is the probability that a dark matter particle belongs to a stream with a density greater than $\rho_\text{s}$. As described above, solving this equation involves a number of unknowns that are beyond the scope of the current paper which is dedicated to describing the new $\Lambda$CDM prediction of concentrated dark matter hairs projecting from compact bodies.  A comprehensive solution for $n_\text{probe}$ would look at a realistic experimental setup and provide likelihood parameter figures drawn from a sample of simulated planetary hair realizations as well as incorporate orbital/fuel constraints.  However, one can get a rough idea of the difficulty  of finding a hair by approximating $\bar{M}\sim M/\bar{h}_\text{w}$, and taking $M\sim10^8$ for Jupiter and $M\sim10^7$ for Earth to obtain
\beq\label{jupnp}
n_\text{probe} &>& \frac{ \langle\rho\rangle }{ \rho_{\text{s}}}  \frac{op}{F(>\rho_{\text{s}})}~~\text{for Jupiter} \\ \label{earnp}
n_\text{probe} &>& 10^4 \frac{ \langle\rho\rangle }{ \rho_{\text{s}}}  \frac{op}{F(>\rho_{\text{s}})}~~\text{for Earth}
\eeq
The reason $\rho_\text{s}$ appears in Eqs.~(\ref{jupnp},\ref{earnp}) is because the number of streams at the smallest densities dwarfs the number of streams with the highest densities.  For example, the simulations of Ref.~\cite{Vogelsberger:2010gd} see $10^{14}$ streams near the Sun but only about $10^6$ are massive.  The way to interpret the above equations is as a lower limit on the number of probes required to see hairs generated from streams with density $\rho_\text{s}$.  With this understanding, it is seen from Fig~10 of Ref.~\cite{Vogelsberger:2010gd} that the ratio $ \langle\rho\rangle /[ \rho_{\text{s}}F(>\!\rho_{\text{s}})]\cong10^5$ for the  densest streams ($\rho_\text{s}/\langle\rho\rangle >10^{-4}$) with the ratio increasing very fast with decreasing $\rho_\text{s}$.  In particular, the least dense streams are effectively invisible because their hair width  $\bar{h}_\text{w}$ is tiny as seen from \Eq{hw}.  For the dense streams with finite $\bar{h}_\text{w}$, $n_\text{probe}$ seems prohibitively high especially since the parameter $o$ can be large.

This rough estimate based on the simulations of Ref.~\cite{Vogelsberger:2010gd} and assuming Earth-based detector sensitivity therefore appears to put  hair detection out of reach, unless 1) $p$ can be constrained to be quite small, 2) the uncertainty on the radial distance to the hair root can be narrowed enough that  density enhancements of $10^8-10^9$ for Earth and $10^{10}-10^{11}$ for Jupiter can be used above, suppressing $n_\text{probe}$ by a factor of $10^2-10^4$, 3) future theoretical/numerical developments  allow the high-$\rho_\text{s}$ slope of Fig~10 from Ref.~\cite{Vogelsberger:2010gd} to be steeper\footnote{Specifically, Monte Carlos simulations should be run using the {\it Planck} cosmological parameters~\citep{Ade:2015xua}, different gravitational softening values and narrower radial shells to estimate error bars on the slope.}  and 4) a way is found to exploit the characteristics of the space environment (such as microgravity)  to design novel direct dark matter detectors, for both WIMPs and axions, with better sensitivity.

One highly developed technology for space missions is refrigeration for experiments requiring very low temperatures typical of axion searches~\citep{Asztalos:2009yp}.  The cosmic microwave background (CMB) {\it Planck} satellite mission relied on active cooling to lower the temperature of the High Frequency Instrument bolometer plate to 93~mk~\citep{Planck:2011aj} which is similar to the temperature goal for the ADMX apparatus upgrade.  Generally speaking, axion searches and CMB space telescopes broadly share a reliance on microwave science and engineering which would make it more likely  that a space-based axion detection mission could recycle well-tested CMB technology.  Low temperature  detectors could also be efficaciously used for a WIMP detection space mission thanks to their excellent signal-to-noise ratio and their event-by-event background rejection capabilities which may reduce the cosmic rays shielding requirements~\citep{Mirabolfathi:2013hda,Guo:2013dt}.  Since discovering a hair is a separate goal with different benefits from  direct  dark matter detection (which may happen on Earth first), it may be possible to transfer improvements in detector sensitivity to a space mission that would greatly increase the odds of finding a hair.

Turning next to whether the compact body hairs could contribute a measurable signal to indirect dark matter searches from dark matter annihilation, consider the usual ambient contribution estimated  from the equation
\beq
\Phi(E_\gamma,\psi) = \frac{\langle\sigma v\rangle}{8\pi m_\chi^2} \delta(E-E_\gamma) \int_{\text{LOS}} \rho^2(l,\psi)dl~,
\eeq
where a monochromatic $\gamma$-ray decay was assumed, $m_\chi$ is the CDM mass and the astrophysical content of the flux $\Phi$ is contained in the line-of-sight integral on the end.  Including the planetary focussing, the astrophysical integral becomes
\beq
& &\int_{\text{LOS}} \rho^2(l,\psi)dl \rightarrow \int_{\text{LOS}} \rho^2(l,\psi)dl + G_{\text{L}}(\phi, \psi) \\
& &G_{\text{L}}(\phi,\psi) =  \int_{\text{LOS}}\text{d}l \Omega_{\text{CO}}(l,\phi,\psi) \nonumber \\
& &~~~~~~~~~~\times \sum_{\text{hair}} P_{\text{hair}}(l,\phi,\psi) \int_{\text{hair}} \text{d}t\text{d}A(t) \rho_{\text{hair}}^2(t,l,\psi)~,
\eeq
where $\Omega_{\text{CO}}(l,\phi,\psi)$ is the density of compact objects in the direction $(\phi,~\psi)$ that focus the CDM beyond their surface, $ \rho_{\text{hair}}(t,~l,~\psi)$ is the CDM density at point $t$ along the hair of the gas giant, a distance $l$ away from us with an angle $\psi$ from the galactic center and $P_{\text{hair}} $ is the probability of seeing a hair with density $\rho_{\text{hair}}$; the last integral is performed about the volume of the hair with d$A$ the area of the cross section at $t$.  It is immediately clear why that second contribution is negligible: $\Omega_{\text{CO}}$ has units of $\text{kpc}^{-2}$  and can be completely ignored in CDM annihilation searches.  This conclusion also holds locally as the number of annihilation products produced by each unit volume within the solar system far exceeds the number  produced along the hairs of planetary bodies even after accounting for the density enhancement.  This result also negates the possibility of detecting dark matter annihilation in hairs surrounding a particular compact body (such as a white dwarf or a neutron star)  with a telescope, as it would be swamped by the ambient dark matter annihilation not to mention the directional suppression stemming from the tiny opening angle  of a created photon propagating from the compact body to an observing satellite orbiting the Earth.  This is also true for solar system bodies like Jupiter and Earth, which cannot produce detectable amounts of annihilation products.

Lastly consider the possibility of  relic neutrino hairs.  The relic neutrino energies follow a Fermi-Dirac distribution and cannot form the caustics necessary to produce hairs, since their velocity distribution is completely smooth with a large dispersion.
\\
\\
{\bf SUMMARY.} A new $\Lambda$CDM prediction of concentrated dark matter hairs extending from compact bodies was described.  The hairs are  caustics of dark matter streams with primordial velocity dispersion passing through compact bodies.  These caustics appear for both massless and massive particles but only the massive particles have caustics located near the lensing planet. It was shown that the existence of  hairs is robust for realistic radial density profiles.  It was also shown that the density enhancements are neither affected by the detailed structure of  the body's radial density profile, nor the dark matter stream velocity. Hair density enhancements were calculated for 3 cases: the analytic constant density Earth, the PREM and the J11-4a density model of Jupiter.  In all cases, large density enhancements were shown to exist up to $10^9$ for Earth and $10^{11}$ for Jupiter.  Furthermore, the density enhancements of the hairs  clearly reflected the boundary layers of the density profiles of the planets.  The discovery of a hair would therefore be a huge boon for both planetary science and cosmology:
\begin{itemize}
\item A hair would provide a uniquely powerful laboratory to study dark matter interactions.
\item Hairs are unique windows into the fine structure of the local dark matter streams.
\item A hair discovered near Earth (or Jupiter) would find counterparts in all solar system bodies providing a powerful universal tool to probe the interior of almost any planet or moon.
\end{itemize}
A  set of parameters determining the feasibility of finding a hair was listed and an approximate equation for the likelihood of discovering a hair was written down, given certain constraints and assumptions.  Implications for indirect dark matter searches and relic neutrinos  were shown to be inexistent.
\\
\\
{\bf ACKNOWLEDGEMENTS.} The author would like to thank Takeyasu Ito, Charles Lawrence and Brad Plaster for useful suggestions and comments.  This research was carried out at the Jet Propulsion Laboratory, California Institute of Technology, under a contract with the National Aeronautics and Space Administration and funded through the internal Research and Technology Development program.  This work was also supported by NASA through the US Planck collaboration. \copyright~2015 California Institute of Technology. Government sponsorship acknowledged.


\bibliography{hair_bib}{}
\end{document}